 \documentclass[aps,pre,twocolumn,amsmath,amsfonts,floatfix,superscriptaddress,nofootinbib,a4paper]{revtex4-2}

\usepackage{mathrsfs} \usepackage{amssymb} \usepackage{graphicx,color} \usepackage{bbm} \usepackage{multirow}
\usepackage{bm} \usepackage{mathrsfs} \usepackage{commath} \usepackage{stmaryrd} \usepackage{color} \usepackage{stmaryrd}
\usepackage{tikz}\usetikzlibrary{shapes,arrows,positioning,calc}\usepackage{comment}\usepackage{dsfont}
\usepackage{hyperref}
\hypersetup{
    colorlinks,
    linkcolor={blue!70!black},
    citecolor={green!70!black},
    urlcolor={red!70!black}
}

\usepackage{tikz}\usetikzlibrary{shapes,arrows,positioning,calc,shapes.misc}
\usepackage{comment}\usepackage{bbold}
\usepackage{esint}\usepackage{upgreek} 
\usepackage{braket} 

\tikzset{cross/.style={cross out, draw=black, minimum size=2*(#1-\pgflinewidth), inner sep=0pt, outer sep=0pt},
cross/.default={1pt}}

\let\a=\alpha \let\b=\beta \let\g=\gamma \let\d=\delta
   \let\k=\kappa
\let\l=\lambda \let\m=\mu \let\n=\nu  \let\p=\pi
\let\s=\sigma \let\t=\tau  
   \let\G=\Gamma
\let\D=\Delta \let\L=\Lambda  
   
 \let\r=\rho \let\th=\theta \let\io=\infty
\let\om=\omega
\def\ie{{\textit{i.e.} }}\def\eg{{\textit{e.g.} }}

\def\PP{{\cal P}} 
\def\FF{{\cal F}} 
 \def\BB{{\cal B}}
\def\LL{{\cal L}}  
\def\DD{{\cal D}} \def\SS{{\cal S}}

\def\Re{{\rm Re}\,}\def\Im{{\rm Im}\,}

\def\to{\rightarrow}  
\def\RRR{\mathbbm{R}}

\def\dd{\mathrm{d}}

\def\Tr{\mathrm{Tr}}
\def\dd{\mathrm{d}}

\def\restriction#1#2{\mathchoice
              {\setbox1\hbox{${\displaystyle #1}_{\scriptstyle #2}$}
              \restrictionaux{#1}{#2}}
              {\setbox1\hbox{${\textstyle #1}_{\scriptstyle #2}$}
              \restrictionaux{#1}{#2}}
              {\setbox1\hbox{${\scriptstyle #1}_{\scriptscriptstyle #2}$}
              \restrictionaux{#1}{#2}}
              {\setbox1\hbox{${\scriptscriptstyle #1}_{\scriptscriptstyle #2}$}
              \restrictionaux{#1}{#2}}}
\def\restrictionaux#1#2{{#1\,\smash{\vrule height .8\ht1 depth .85\dp1}}_{\,#2}}

\newcommand{\beq}{\begin{equation}} \newcommand{\eeq}{\end{equation}}
 \newcommand{\wt}{\widetilde}

\newcommand{\argc}[1]{\left[#1\right]}
\newcommand{\arga}[1]{\left\lbrace #1\right\rbrace }
\newcommand{\argp}[1]{\left(#1\right)}

\newcommand{\moy}[1]{\left\langle  #1 \right\rangle }

\def\mw{\omega_{\rm MW}}
\def\ome{\om_{\rm e}}

\def\ketbra#1#2{{\vert#1\rangle\!\langle#2\vert}}

\def\sign{\textrm{sign}}

\def\SI{See the Supplementary Material [SM] for further details.}
\newcommand{\SIcite}[1]{[SM,#1]}

\begin{document}

\title{Bath-induced Zeno localization in driven many-body quantum systems} 

\author{Thibaud Maimbourg}
\email{thibaud.maimbourg@lptms.u-psud.fr}
\affiliation{LPTMS, CNRS, Universit\'e Paris-Saclay, 91405, Orsay, France.}

\author{Denis M. Basko}
\affiliation{Universit\'e  Grenoble  Alpes  and  CNRS,  LPMMC,  25  rue  des  Martyrs,  38042  Grenoble,  France}

\author{Markus Holzmann}
\affiliation{Universit\'e  Grenoble  Alpes  and  CNRS,  LPMMC,  25  rue  des  Martyrs,  38042  Grenoble,  France}

\author{Alberto Rosso}
\affiliation{LPTMS, CNRS, Universit\'e Paris-Saclay, 91405, Orsay, France.}

\begin{abstract}
We study a quantum interacting spin system subject to an external drive and coupled to a thermal bath of 
vibrational modes, uncorrelated for different spins,
serving as a model for dynamic nuclear polarization protocols. We show that even when the many-body eigenstates 
of the system are ergodic, a sufficiently strong coupling to the bath may effectively localize the spins 
due to many-body quantum Zeno effect. 
Our results provide an explanation of the breakdown of the thermal mixing regime experimentally observed above 4 -- 5 Kelvin in these protocols.
\end{abstract}

\maketitle


The thermalization of an isolated many-body quantum system stems from two mechanisms. The first is dephasing, \ie the projection by 
the unitary dynamics of the initial state onto the Hamiltonian eigenstates. The second is the matching between the 
expectation value of physical observables in these eigenstates and those of the microcanonical ensemble. 
This second property, called the eigenstate thermalization hypothesis (ETH)~\cite{LL80, D91, S94, DAKPR16}, 
implies that if the quantum state is a mixture of Hamiltonian's eigenstates, the system appears thermal.
Consequently, one expects that even in the presence of a drive and dissipation, a unique effective temperature characterizes the stationary state~\cite{DLR15,DLRAMR16,LAR18,LARA19,LLR17}.

 A celebrated confirmation of this scenario is the  {\em thermal mixing}  reached in dynamic nuclear polarization (DNP)~\cite{abragam,DLR15}, a 
  protocol used for NMR applications. A sample, doped with molecules possessing unpaired electron spins, is exposed 
  to a strong magnetic field, frozen at temperature   $\b^{-1} \sim 1$ K and driven out of equilibrium by microwave 
  irradiation  at frequency $\mw$. After one hour all nuclear species in the sample  ($^{1}H$, $^{13}C$, $^{15}N$...) 
  thermalize to a single temperature $\beta_s^{-1}$, called \textit{spin temperature}~\cite{abragam,GMRABIK17}. By tuning $\mw$, one can reach $\b_s\gg\b$ 
  which strongly hyperpolarizes the nuclear spins, an essential aim in NMR spectrometry and imaging. 
Inconveniently, this regime disappears above $4-5$~K and nuclear polarization becomes weak~\cite{SWS94,GK08,HKSFGV15}.

In this Letter, we show how a coupling to a local bath can explain the thermal mixing breakdown and reveal fingerprints of 
localization in this nonequilibrium steady state. 
Such ergodicity breaking is not caused by a violation of ETH, 
as in strongly disordered systems (a phenomenon called many-body localization, MBL~\cite{AGKL97,BAA06,NH15,AL18,AABS19}), but by a competition between dephasing and 
system-bath interaction that prevents the stationary state from being an eigenstate mixture. 
This phenomenon is a many-body analog of the quantum Zeno effect~\cite{BN67,K68,MS77,IHBW90,FP01,BP02,HNPRS15}, 
where infinitely frequent measurements impede the unitary evolution of a single degree of freedom. Here the interaction with bath modes, uncorrelated for different spins, plays the role of the measurements.   
We show that going beyond the traditional scheme of weak coupling to the 
bath~\cite{DLR15,DLRAMR16,RAMRDL18,LAR18,LARA19,LLR17}, through a recently proposed approach not relying on the secular approximation~\cite{NR20,KFW18,KSKS20}, is necessary to account for this type of localization.

  \begin{figure}[t]
 \includegraphics[width=\linewidth]{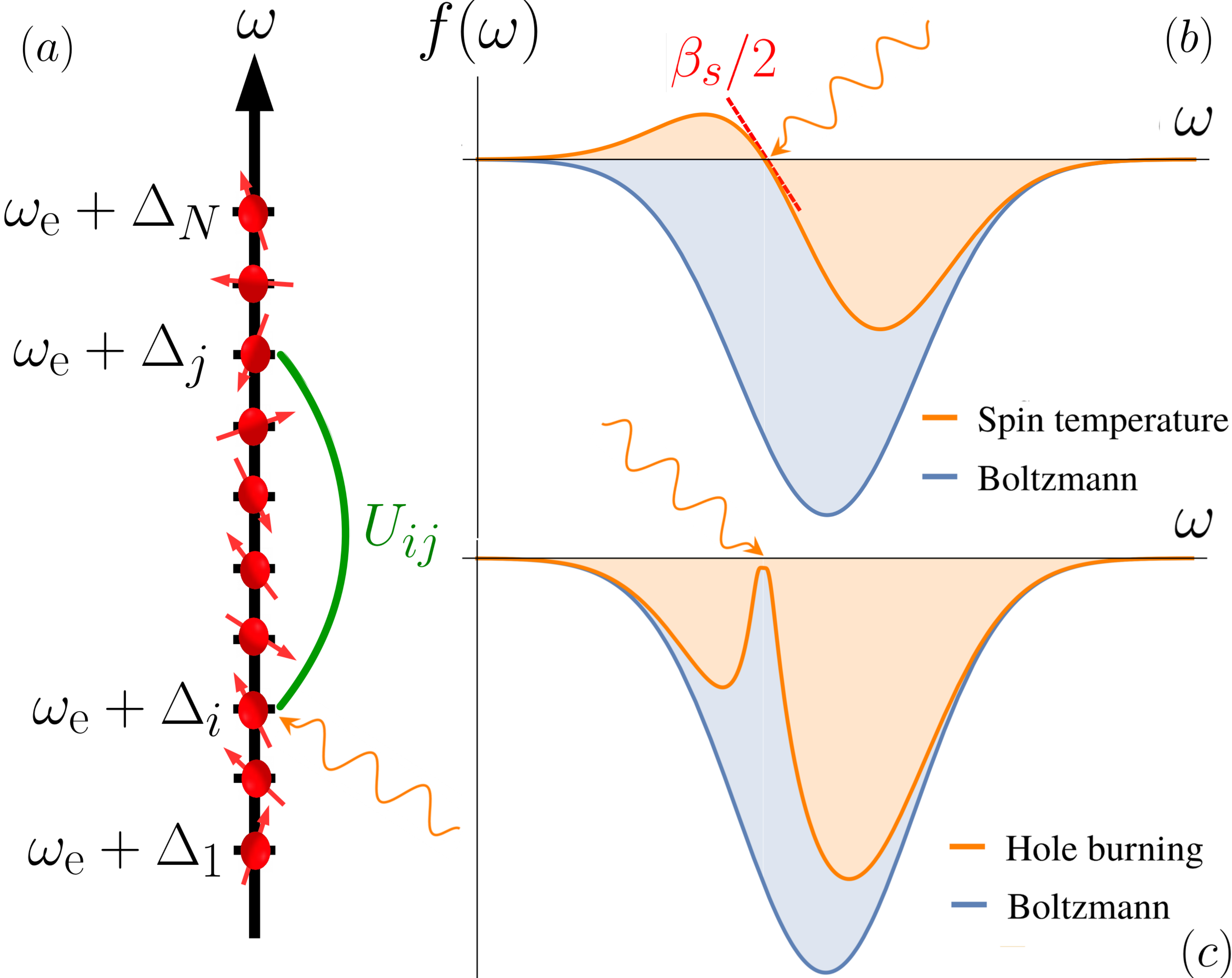}
 \caption{(a) Sketch of the system: $N$ electron spins with dipolar interactions of strength $U_{ij}$ in a strong inhomogeneous magnetic field $\ome+\D_i$
 in contact with a thermostat are irradiated by microwaves at frequency $\mw$ (wavy arrow). 
 (b)(c) EPR spectrum $f(\om)$ (defined via Eq.~\eqref{eq:polim}) at Boltzmann equilibrium (blue curve) and under microwave driving 
 displaying two shapes (orange curves): (b) the spin-temperature 
profile~\cite{A70,abragam} with linear behavior (dashed red) of slope $\b_s/2$ close to resonance $\om=\mw$ 
(c) the hole burning at resonance.}
 \label{fig:schema}
\end{figure}

The DNP arises from the steady state of $N$ unpaired electron spins. 
Their Hamiltonian reads (Fig.~\ref{fig:schema}(a))
\begin{equation}\label{eq:HS}
  \hat H_{\rm S}=\sum_{i=1}^N (\om_{\rm e}+\D_i)\hat S_i^z+\hat H_{\rm dip} \ .
\end{equation}
$\ome$ is the strong magnetic field along the $z$ axis (Zeeman gap). 
$\D_i$ is a small disorder from the random orientation of the molecule where the spin lies and $\hat H_{\rm dip}$ 
stands for the dipolar interaction. The large 
magnetic field implies $\hat S^z=\sum_i\hat S_i^z$ is 
conserved, hence the dipolar Hamiltonian gets truncated as~\cite{abragam,AG78,SPG92,DLRAMR16}
\begin{equation}\label{eq:dip}
 \hat H_{\rm dip}=\sum_{i<j} U_{ij}\argp{ \hat S_i^+\hat S_j^-+\hat S_i^-\hat S_j^+-4\hat S_i^z\hat S_j^z} \ ,
\end{equation}
where $U_{ij}$ depends on the distance between the spins and their orientation with respect to the magnetic field. The spins are in contact with a thermal bath and driven by microwaves through $\hat H_{\rm MW}(t)=\om_1\argc{\hat S^x \cos(\mw t)+\hat S^y \sin(\mw t)}$.
When $\mw\approx\ome$, the electron spins reach a 
stationary state 
 probed experimentally by measuring the electron paramagnetic resonance (EPR) spectrum. Two typical shapes can occur: $\textit{(i)}$ a linear  curve close to 
irradiation frequency (Fig.~\ref{fig:schema}(b)). Electrons are in thermal mixing, \ie equilibrated at the spin temperature 
$\b_s^{-1}$ and interacting via 
$\hat H_{\rm S}$ with a shifted magnetic field $\ome\to\ome-h$ where $h\simeq\mw$. Through hyperfine interaction with the electron spins, 
the nuclear spins thermalize with polarization $P_{\rm n}=\tanh(\b_s\om_{\rm n}/2)$ ($\om_{\rm n}$ is the nuclear Zeeman gap)~\cite{abragam,DLR15,DLRAMR16}. 
$\textit{(ii)}$ a ``hole burning'' close to irradiation (Fig.~\ref{fig:schema}(c)) found by Bloch~\cite{B46} for noninteracting spins%
: the resonant spins ($\ome+\D_i\simeq\mw$) are brought to a high 
temperature while off-resonant ones remain at $\b^{-1}$. 
The hole-burning shape is recovered in the MBL regime~\cite{DLR15,DLRAMR16,RAMRDL18,LAR18,LARA19}, 
revealing different local temperatures: the nuclear species exhibit weak polarization not accounted for by a single temperature.

The spin-temperature shape was observed long ago in the EPR spectrum of $\textrm{Ce}^{3+}$ in 
$\textrm{CaWO}_4$ crystals~\cite{A70}. 
More recently, experiments on irradiated EPR spectrum retrieved instead a hole-burning shape~\cite{SWS94,GK08,HKSFGV15}
above $4-5$~K.  
In the following, we argue that even in ETH systems, a bath of uncorrelated modes triggers 
quantum jumps 
that can induce localization \textit{if} bath transitions prevail over dephasing. 
In particular, we compute the EPR spectrum and show a crossover from a spin-temperature to a hole-burning shape. 
We interpret this crossover as a manifestation of a bath-induced Zeno localization in the many-body 
eigenbasis of $\{\hat{S}_i^z\}$. 
As bath transitions become more effective when raising temperature, curiously, this ergodicity breaking happens in the high-temperature phase.

\paragraph*{Effective dynamics due to the bath --}
The electron spins are dilute, so we assume they are in contact with vibration modes $\hat B_i^\m$~\SIcite{Sec.V}~\footnote{\SI} that are uncorrelated (thus they cannot induce effective interactions between spins):
\begin{equation}
 \hat H_{\rm int}=\sum_{\substack{i=1,\dots,N\\\m=x,y,z}} \hat S_i^\m \otimes \hat B_i^\m \ .
\end{equation}
Assuming the bath is equilibrated at temperature $\b^{-1}$, we trace out the $\hat B_i^\m$ variables
in the full density matrix $\r_{\textrm{S}\otimes\textrm{B}}$ and write an effective evolution for the spin system density matrix 
$\r=\Tr_{\rm B}(\r_{\textrm{S}\otimes\textrm{B}})$. The ensuing evolution is no longer unitary but must still preserve the trace and semipositivity 
of $\r$. The most general Markovian dynamics must then be of the 
Gorini-Kossakowski-Sudarshan-Lindblad (GKSL) form~\cite{BP02}:
\begin{equation}\label{eq:GKSL}
 \dot\r=-i[\hat H,\r]+\sum_{\a}\hat A_\a\r \hat A_\a^\dagger-\frac12\arga{\hat A_\a^\dagger\hat  A_\a,\r } \ ,
\end{equation}
where $\hat H$ is Hermitian and $\{\hat A_\a\}$ is a set of jump operators. 
To integrate the bath degrees of freedom~\SIcite{Sec.I}, we consider weak spin-bath coupling and perform a 
perturbative expansion of the full unitary dynamics of $\r_{\textrm{S}\otimes\textrm{B}}$ at second order in 
$\hat H_{\rm int}$. The Born-Markov approximation~\cite{Al89,AGKK12,BP02}  
yields an effective Markovian evolution for $\r$. 
The uncorrelated bath degrees of freedom are described by a single equilibrium correlation function
\begin{equation}\label{eq:rate}
 {\g(\om)=\int_{-\io}^\io\dd \t\,e^{i\om\t}\moy{\hat B_i^\m(\t)\hat B_i^\m(0)}_{\rm B}} =\frac{h(\om)}{T(|\om|)}\ ,
\end{equation}
where $h(\om)=(1+e^{-\b\om})^{-1}$ enforces detailed balance, 
while $T(|\om|)$ is the timescale of energy exchange $\om$ with the spins.

The Markovian approximation is not unique and in general not in GKSL form~\eqref{eq:GKSL}.
We implement the Markovian prescription of~\cite{NR20,KFW18,KSKS20}, which leads to a GKSL form setting $\hat H = \hat H_{\rm S}$ 
and 
\begin{equation}\label{eq:nonsecjump}
\hat A_\a = \sum_{n,m}\sqrt{\g(\om_{nm})}\bra{m}\hat S_i^\m\ket{n} \ketbra{m}{n} \ ,
\end{equation}
with $\a=(i,\m)$ and $\om_{nm}=\varepsilon_n-\varepsilon_m$ are $\hat H_{\rm S}$ energy gaps. 
We have three typical timescales $T(|\om|)$:
\textit{(i)} $T_1$  for transitions of energy gap $\pm\ome$, giving the jump operators
\begin{equation}\label{eq:simpnonsecxy}
 \begin{split}
    \hat A_i^x=& \sqrt{\frac{h(\ome)}{2T_1}}\argp{\hat S_i^-+e^{-\b\ome/2}\hat S_i^+}\ ,\\
  \hat A_i^y=& i\sqrt{\frac{h(\ome)}{2T_1}}\argp{\hat S_i^--e^{-\b\ome/2}\hat S_i^+}
 \end{split}
\end{equation}
\textit{(ii)} $T^*$ for transitions of finite energy $|\om|\ll\ome$, and
\textit{(iii)} $T(0)$ for zero-energy  transitions within the same eigenstate, giving
\begin{equation}\label{eq:simpnonsecz}
 \hat A_i^z= \frac{\hat S_i^z}{\sqrt{2T^*}}+ \argp{\frac{1}{\sqrt{2T(0)}}-\frac{1}{\sqrt{2T^*}}} \sum_n \bra{n}\hat S_i^z\ket{n}\ketbra{n}{n}\ .
\end{equation}
The \textit{nonsecular} jump operators~\eqref{eq:simpnonsecxy},\eqref{eq:simpnonsecz} (when $T(0)\approx T^*$) are well localized in space.
The quantum trajectories result from a competition between the unitary dynamics 
projecting on thermal eigenstates 
and repeated measurements performed by the nonsecular jump operators. 
If jump rates dominate, thermalization is  hampered
in a way reminiscent of the quantum Zeno effect. 

This choice of jump operators contrasts with the usual 
weak-coupling prescription~\cite{Al89,AGKK12,BP02}, where 
a GKSL equation~\eqref{eq:GKSL} is recovered through an additional \textit{secular} approximation. The jump operators select only a given transition 
energy $\om_{nm}$ between eigenstates $\ket m$ and $\ket n$: 
\begin{equation}\label{eq:jumpsec}
 \hat A_\a^{\rm sec}(\om_{nm}) = \sqrt{\g(\om_{nm})}\bra{m}\hat S_i^\m\ket{n} \ketbra{m}{n}\ .
\end{equation}
The secular jump operators~\eqref{eq:jumpsec} are directly projected on the eigenstates of $\hat{H}_\mathrm{S}$,  unlike jumps~\eqref{eq:simpnonsecxy},\eqref{eq:simpnonsecz} for which the secular approximation is released. 
With these nonsecular jumps, the system's state gets projected on the Hamiltonian eigenstates only if bath timescales $T_1$, $T^*$ are long with respect to dephasing. 
To emphasize the effect of the secular approximation, we insert Eq.~\eqref{eq:jumpsec} into Eq.~\eqref{eq:GKSL},
yielding an exponential decay of the off-diagonal elements (coherences) in the eigenbasis:
\begin{equation}\label{eq:secproj2}
  \dot\r_{nm}= -\argp{i\om_{nm}+\frac{1}{T_{nm}}}\r_{nm} \ ,\\
\end{equation}
where $i\om_{nm}$ is the dephasing due to the unitary evolution and $T_{nm}>0$ is the decoherence~\SIcite{Sec.I.B}  
originating from the bath timescales. Therefore one can work in the \textit{Hilbert approximation} where the dynamics is projected on 
the diagonal elements: it amounts to transit from an eigenstate to the other 
with rates given in~\SIcite{Sec.I.E}. The nonsecular dynamics~\eqref{eq:nonsecjump} adds to the right-hand side 
of Eq.~\eqref{eq:secproj2}  entries other than $\r_{nm}$, with associated bath rates~\SIcite{Sec.I.F}, allowing the existence of steady-state coherences. 

\paragraph*{Microwave drive --}
In equilibrium, the steady state is in practice accurately described by the Boltzmann distribution 
with either choice of jump operators~\cite{NR20,LY20comment,NR20response}. The nonsecular evolution brings drastic changes out of equilibrium: the drive creates 
an imbalance that probes localization. In a DNP protocol the system is irradiated by microwaves 
described by $\hat H_{\rm MW}(t)$. 
They induce local temperature inhomogeneities as resonant spins get hotter while others are frozen by the low-temperature bath. 
The dynamics of the rotating-frame density matrix 
$e^{i\mw t \hat S^z}\r(t)e^{-i\mw t \hat S^z}\to \r(t)$ remains given by Eq.~\eqref{eq:GKSL} with the 
shift $\hat H=\hat H_{\rm S}-\mw \hat S^z+\om_1\hat S^x$~\SIcite{Sec.I.D}. 

\begin{figure}[t]
\centering
\includegraphics[width=\linewidth]{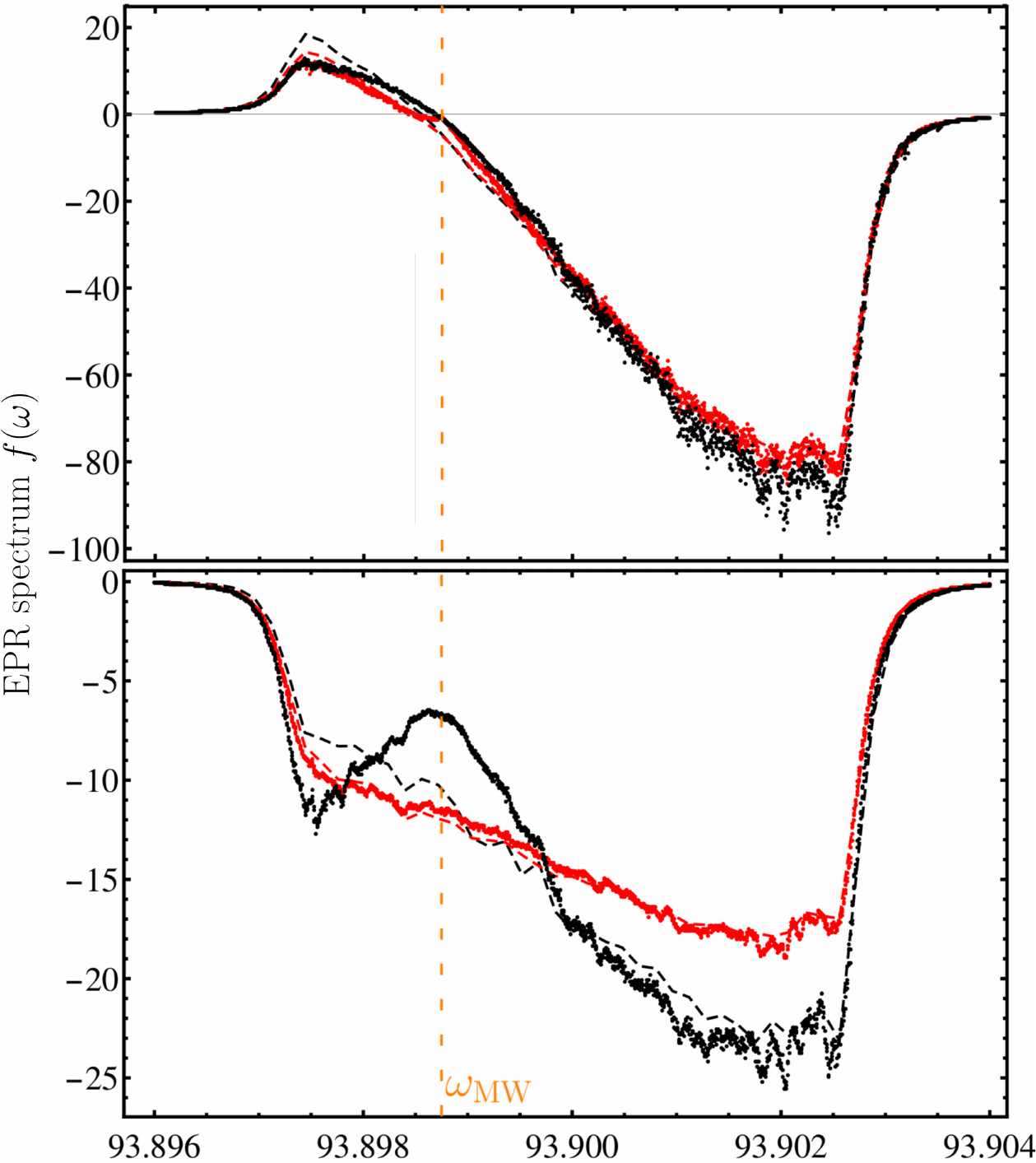}\\
$\om$ $(2\p$GHz)
\caption{\label{fig:EPR} EPR spectra :
dots represent numerical profiles for Hilbert (red) and nonsecular (black) evolutions.
Dashed lines are calculated through a spin-temperature ansatz.
(top) $\b^{-1}=1.2$ K. 
The bath is slow, nonsecular and Hilbert dynamics provide analogous spin-temperature curves.
(bottom) $\b^{-1}=12$ K.  Bath timescales are short, 
the nonsecular dynamics gets localized and displays a hole burning. 
Here the Hilbert approximation fails, predicting a spin-temperature behavior.
Averages are done over 1000 realizations.}
\end{figure}

\paragraph*{Numerical computation of the EPR spectrum --} 
We compare the stationary states predicted by the Hilbert dynamics with the ones obtained 
by the nonsecular evolution Eq.~\eqref{eq:GKSL} with jumps~\eqref{eq:simpnonsecxy},\eqref{eq:simpnonsecz}
for uniform disorder $\D_i \in [-\frac{\D\ome}2,\frac{\D\ome}2]$ and $U_{ij}$ mimicked by 
independent Gaussian distributions with zero mean and variance $U^2/N$.
We fix $\D\ome=5\cdot2\p$MHz, $U=0.75\cdot2\p$MHz (note that $\ome=93.9\cdot2\p$GHz)
where $\hat H_{\rm S}$ has ETH statistics. 
We consider two temperatures; at high temperature the bath timescales 
are short (Table~\ref{tab:times})[SM,Sec.V]. 

\begin{table}[h!]
\begin{center}
  \begin{tabular}{c| c | c | c | c | c }
     $\b^{-1}$(K) & $T(0)$($\m$s) & $T^*$($\m$s) & $T_1$($\m$s) & $\om_1$($2\p$MHz)  & $\mw$($2\p$GHz)\\
     \hline
   1.2  &  1.6    &  $80$ & $160$ & 0.628 & 93.8988   \\
	\hline
	12  &  $0.16$  &  $0.16$  & $1.6$ & 0.628  & 93.8988
  \end{tabular}
  \end{center}
    \caption{Control parameters chosen for the system at two temperatures, close to experimental 
    values~\cite{HFV13,HKSFGV15,DLR15}. }
  \label{tab:times}
\end{table}

We compute numerically the steady-state density matrix $\r_{\rm stat}$~\cite{julia}. 
The Hilbert case amounts to a $2^N\times 2^N$ linear system, which for $N=10$ spins 
is treated by exact diagonalization. 
The nonsecular dynamics Eq.~\eqref{eq:GKSL} is instead a $4^N\times 4^N$ linear system requiring  Krylov subspace
methods (biconjugate gradient-stabilized algorithm)~\cite{S03}. 
To probe the stationary state we focus on 
the EPR spectrum:
starting at time $\t=0$, a $\p/2$ microwave pulse projects the steady-state polarization of a given spin $i$ on the $y$ axis,   
$\r_{\p/2}=e^{i\frac\p 2\hat S_i^x}\r_{\rm stat}e^{-i\frac\p 2\hat S_i^x}$. For short times after the pulse, the evolution is unitary 
and the polarization in the $(x,y)$ plane is encoded in 
\begin{equation}\label{eq:polim}
 g_i(\t)=
 -2i\Tr\argc{\hat S_i^+(\t)\r_{\p/2}} \ ,
\end{equation}
where $\hat S_i^+(\t)=e^{i\hat H_{\rm S}\t} \hat S_i^+ e^{-i\hat H_{\rm S}\t}$. 
The EPR spectrum is defined by Fourier transform:
$f(\om)=\frac 1 N\sum_i\Re\argc{\int_0^\io\frac{\dd \t}{\p}\,g_i(\t)e^{-i\om\t}}$~\SIcite{Sec.II}. 

At low temperature, we observe a spin-temperature curve for both dynamics in the EPR spectra (Fig.~\ref{fig:EPR}).  
Here the bath timescales $T^*$, $T_1$ are long with respect to both dephasing time $\om_{nm}^{-1}\approx \min(1/U,1/\D\ome)$ and decoherence time $T_{nm}\approx T(0)$. Consequently, the density matrix gets projected 
in the eigenstate basis, as in the Hilbert approximation.
At higher temperature, the EPR spectrum is spin-temperature-like for Hilbert dynamics, whereas it has
a hole-burning shape in the nonsecular evolution. 
The spin-temperature behavior observed in the Hilbert approximation is expected~\cite{DLR15,DLRAMR16,RAMRDL18}:
due to ETH, the jump operators projected on eigenstates generate 
energy and polarization changes without any other information such as the spatial location of the spins. 
Conversely, in the nonsecular equation the jump operators are well localized in space and 
compete with dephasing, which is unable to project the system on the eigenstates (as $T^*$ becomes comparable to dephasing and decoherence times).  
The EPR spectrum develops a hole burning similar to the one already observed when $\hat H_{\rm S}$ has MBL eigenstates~\cite{DLR15,DLRAMR16}\SIcite{Fig.S3}, although $\hat{H}_\mathrm{S}$ eigenstates are ergodic for our parameters. 
This breakdown is confirmed by comparing the EPR profiles with the ones (dashed lines in Fig.~\ref{fig:EPR}) 
obtained through a spin-temperature ansatz for the steady-state 
density matrix $ \r_{nn}^{\rm ans}(\b_s,h)\propto e^{-\b_s(\varepsilon_n-hs_n^z)}$. $s^z_n$ are  eigenvalues of the conserved 
$\hat S^z$. The spin temperature $\b_s^{-1}$ (respectively magnetic field $h$) is 
conjugated to the energy (respectively polarization) and determined by a fit \cite{DLRAMR16}\SIcite{Sec.III.B}. 

\begin{figure}[t]
\centering
\includegraphics[width=\linewidth]{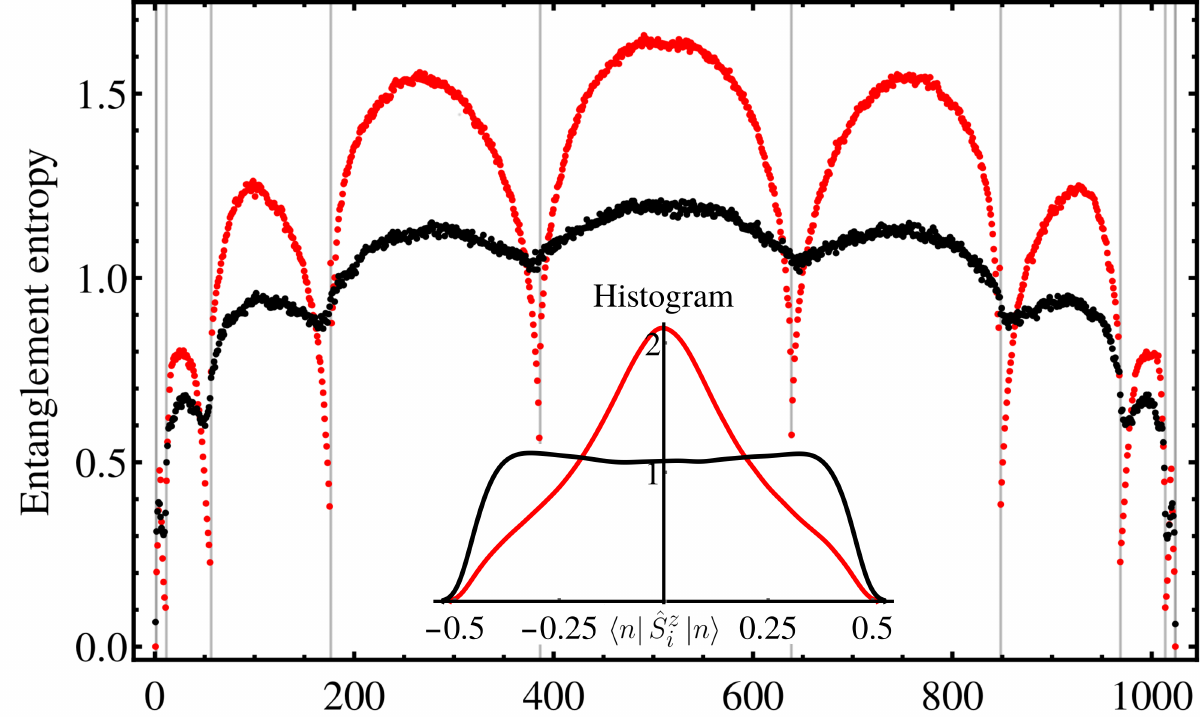}\\ 
$\qquad n$
\caption{\label{fig:EE} 
(main) Entanglement entropy of $\r_{\rm stat}$'s eigenstates
(red : Hilbert, black : nonsecular) 
classified by increasing energy, 
computed through partial trace of each eigenstate $\ket n\bra n$ over spins 5 to 10. Vertical lines delimit polarization 
sectors. 
(inset) Smoothed histogram of $S_i^z$ expectation values ($i=1,\dots,N$) for 100 eigenstates in the middle 
of the $s^z_n=0$ sector. 
For the Hamiltonian eigenstates, the observable is peaked around $S_i^z\simeq0$ as required by ETH. 
The nonsecular eigenstates present instead strong fluctuations, as for MBL states.
Parameters are those of Fig.~\ref{fig:EPR}(bottom).}
\end{figure}

\paragraph*{Spectral properties --} 
The localization phenomenon exhibited by the experimentally relevant EPR spectra is revealed through other observables, 
e.g. polarization profiles~\SIcite{Sec.III}. In Fig.~\ref{fig:EE}, we 
compare the entanglement entropy of the $\r_{\rm stat}$ eigenstates  in the Hilbert and nonsecular cases. The latter case is much less entangled and similar to MBL eigenstates~\SIcite{Sec.III.D; Fig.S7 shows a fully localized case}. 
The present scenario is akin to the measurement-induced entanglement transition 
in schematic models 
such as quantum circuits~\cite{LCF18,ANPS19,SRN19,SRS19,TFD20,JYVL20,ZGWGHP20,BCA20,ZRKCMR20}, free fermionic chains~\cite{CTDL19} and interacting bosonic chains~\cite{TZ20,GD20,FA20}.

\paragraph*{Discussion and conclusion --} 
In open systems the external bath permits thermalization even for strong disorder, when eigenstates are localized. 
This is manifested by phonon-induced hopping transport
~\cite{PS91}: the bath supplies or absorbs the energy needed to hop between localized states. 
Here, we have shown how coupling to uncorrelated thermal vibrations can instead induce localization in 
quantum many-body systems with ergodic eigenstates, revealed in presence of a drive. 
Whether this is a sharp transition or a smooth crossover in the thermodynamic limit remains an intriguing open question. 
Going beyond the conventional secular approximation~\cite{BP02} is required to capture this phenomenon. 
The mechanism underlying this many-body Zeno effect is distinct from the Zeno effect in quantum gases 
with localized particle losses~\cite{SK10,B11,ZKBO12,FCKD19} or dephasing~\cite{DMSD20}, where the combined impact of 
local single-particle losses and interactions renormalizes single-particle quantities.
It also differs from localization by subohmic baths at zero temperature~\cite{BM82,Ch82,Sc83,LCDFGZ87}, 
a polaronic effect wiped out by interactions or temperature~\cite{CGLLSS02}.
In the DNP context, our work,
based on heuristic values of the microscopic timescales, provides an explanation for 
the thermal-mixing breakdown upon increasing temperature, 
arising from enhanced dynamics of the vibrational modes. 
The present analysis calls for a thorough experimental test of temperature influence on the 
different hyperpolarization regimes.

%
%

\paragraph*{Acknowledgments --} 
We thank Fabien Alet, Leticia Cugliandolo, Andrea De Luca, Laura Foini, Nicolas Laflorencie, Leonardo Mazza and Marco Schir\'o 
for valuable discussions around different facets of this work. We also are grateful to Filippo Vicentini for advices on the numerical 
methods. We acknowledge funding from the ANR-16-CE30-0023-01 (THERMOLOC) grant and the Galileo Galilei Institute for hospitality during part of this project.


\clearpage

\begin{widetext}

\centerline{\bf \large Supplemental Material}

\vspace*{2cm}

In the supplemental material we provide some technical details and additional numerics. 
In Sec.~\ref{sec:open} we present the derivation of the different master equations analyzed in the main text (secular, nonsecular 
and Hilbert master equations). More precisely, Secs.~\ref{sub:Born-Markov}, \ref{sec:GKSL}, \ref{sub:NRrev}, \ref{sub:Hilb}  pedagogically review existing results for the reader's convenience, whereas Secs. \ref{sub:NRLS}, \ref{sec:MW}, \ref{sub:comparison} detail or simplify the above equations to the case of our DNP system. 
In Sec.~\ref{sec:EPR} we derive a numerically convenient formula to compute the EPR spectrum. 
In Sec.~\ref{sec:numerics} we display additional plots of EPR and polarization profiles, for an ETH and a MBL Hamiltonian. We explain in Sec.~\ref{sub:fit} the fitting 
procedure to a spin-temperature ansatz, first introduced in Ref.~\cite{DLRAMR16},  
and exhibit in Sec.~\ref{sub:spectral}  complementary data concerning the spectral properties of the bath-induced localized steady state .  We also discuss the effect of the different bath timescales in the DNP model, giving a qualitative recipe to get thermal or non-thermal behavior. 
The numerics are done for the $N=10$ spins setup of the main text, but for completeness we analyze as well in Secs.~\ref{sec:HB}, \ref{sub:spectral} a case of full localization with $N=12$ spins. 
In Sec.~\ref{sec:3levels} we study analytically a minimal model consisting of three levels that presents the essential physics examined in this work, that may be helpful in distinguishing different physical regimes and approximations without the additional complications contained in the complex many-body DNP system. 
Finally in Sec.~\ref{sec:bath} we explain the choice of our DNP Hamiltonian and particularly discuss
the coupling to the bath modes. Spatially-localized vibrations are suitable candidates for such a coupling, but we also show how coupling to delocalized bath modes could also be appropriate, if the system (electrons in DNP) is dilute enough in the medium hosting the bath modes. 
A simple model of ballistic phonons allows to compute the microscopic bath timescales $T(|\om|)$ as a function of frequency and temperature, based on known interactions with ballistic phonons (such as direct and Raman processes). The resulting timescales are decreasing with temperature, an essential feature for the Zeno localization discussed in the main text. \\
Note that units are such that $\hbar=1$, $k_B=1$.

\vspace*{2cm}

\tableofcontents
\makeatletter
\let\toc@pre\relax
\let\toc@post\relax
\makeatother

\clearpage

\renewcommand{\theequation}{S\arabic{equation}}
\renewcommand{\thefigure}{S\arabic{figure}}
\setcounter{equation}{0}

\section{Open and driven system: Weak-coupling nonsecular master equation}\label{sec:open}

The electronic spin system is in contact with a thermal reservoir and irradiated by microwaves. The total Hamiltonian reads
\begin{equation}
 H_{\rm tot}=H_{\rm S}+H_{\rm MW}(t)+H_{\rm int}+H_{\rm B}
\end{equation}
In this section we will carefully study the system-bath interaction to derive an effective evolution of 
the spin system. We thus switch the microwaves off $H_{\rm MW}(t)=0$ and shall reinstate them in Sec.~\ref{sec:MW}. The term $H_{\rm int}$ describes the coupling of the bath to the spins and $H_{\rm B}$ the Hamiltonian of the bath alone, assumed to remain in equilibrium at inverse temperature $\b$. The system-bath interaction is written as
\begin{equation}\label{eq:Hint}
  H_{\rm int}=\sum_{\substack{i=1,\dots,N\\\m=x,y,z}}  S_i^\m \otimes  B_i^\m
\end{equation}
where $ B_i^\m$ are Hermitian operators acting on the bath's Hilbert space, representing 
vibrational modes of the glassy medium. We consider the bath degrees of freedom as \textit{uncorrelated} (see Sec.~\ref{sec:bath} for a discussion):
\begin{equation}
\moy{ B_i^\m(t) B_j^\n(t')}_{\rm B} =0 \quad \text{if} \; (i,\m)\neq (j,\n)
\end{equation}
where $\moy{\bullet}_{\rm B}=\Tr(\bullet\,\r_{\rm B})$ are bath averages with $\r_{\rm B}=e^{-\b H_{\rm B}}/\Tr\, e^{-\b H_{\rm B}}$, and $B_i^\m(t)=e^{it H_{\rm B}}B_i^\m e^{-it H_{\rm B}}$.
Besides the bath degrees of freedom are fluctuations that satisfy $\moy{B_i^\m}_{\rm B}=0$.

In the following we first summarize the steps in~\cite[Sec.3.3]{BP02} to treat perturbatively the effect of the bath on 
the system evolution. We then review the standard secular approximation that allow to turn this expansion into a 
Gorini-Kossakowski-Sudarshan-Lindblad master equation. Next we summarize the steps that provide a GKLS form without throwing away nonsecular 
terms, introduced in Ref.~\cite{NR20}, by modifying the Markov approximation. 
Finally we show how to include the microwave drive in this approach and give the main formal difference between secular and nonsecular master 
equations that account for the effects analyzed in this article.

\subsection{Born-Markov approximation at weak coupling}\label{sub:Born-Markov}

We start defining the non-interacting Hamiltonian $H_0=H_{\rm S}+H_{\rm B}$ and the corresponding unitary evolution operator $U_0(t)=e^{-itH_0}$. The evolution of the whole system's density matrix is ruled 
by the Liouville-von Neumann equation which reads $\dot \r_{\rm tot}=-i[H,\r_{\rm tot}]$ in Schr\"odinger's picture. Since we treat $H_{\rm int}$ as a perturbation it is more convenient to start from the interaction picture version of it,
\begin{equation}\label{eq:vN}
 \dot \r(t)=-i[\wt H_{\rm int}(t),\r(t)]
\end{equation}
where by definition $\r(t)=U_0^\dagger(t)\r_{\rm tot}(t)U_0(t)$ and $\wt H_{\rm int}(t)=U_0^\dagger(t)H_{\rm int}U_0(t)$. 

We aim at getting an equation for the reduced density matrix describing the spin system $\r_{\rm S}=\Tr_{\rm B}\r$. 
At weak coupling\footnote{It can be shown through the Nakajima-Zwanzig projection operator formalism~\cite[Sec.9.1]{BP02} 
that Eq.~\eqref{eq:quasiRedfield} is actually exact at second order in $H_{\rm int}$ 
(provided Eq.~\eqref{eq:initL} and a factorized initial condition). Alternatively, a simpler cumulant expansion has been devised by Alicki and collaborators in~\cite{Al89,AGKK12} for the evolution operator $\L$ such that $\r_{\rm S}(t)=\Tr_{\rm B}\r(t)=\L(t,0)\r_{\rm S}(0)$. Taking the time derivative of the latter equation, one recovers the same result at second order in $H_{\rm int}$. This result is guessed in the present perturbative self-consistent derivation.} one \textit{(i)} solves perturbatively Eq.~\eqref{eq:vN} by integrating it once, $\r(t)=\rho(0) -i \int_0^t\dd s\, [\wt H_{\rm int}(s),\r(s)] $, and plugging the solution back in Eq.~\eqref{eq:vN} -- note that there is no approximation made in this step -- \textit{(ii)} makes the \textit{Born approximation} 
$\r(t)\simeq \r_{\rm S}(t)\otimes \r_{\rm B}$. 
One obtains after tracing over the bath
\begin{equation}\label{eq:quasiRedfield}
 \dot \r_{\rm S}(t)=-\int_0^t\dd s\, \Tr_{\rm B}\argc{\wt H_{\rm int}(t),[\wt H_{\rm int}(s),\r_{\rm S}(s)\otimes \r_{\rm B}]}
 \underset{\t=t-s}{=}-\int_0^t\dd \t\, \Tr_{\rm B}\argc{\wt H_{\rm int}(t),[\wt H_{\rm int}(t-\t),\r_{\rm S}(t-\t)\otimes \r_{\rm B}]}
\end{equation}
The latter equation is not affected by the additional term coming from the initial condition when integrating Eq.~\eqref{eq:vN}, as with $ S_i^\m(t)=e^{itH_{\rm S}} S_i^\m e^{-itH_{\rm S}}$,
\begin{equation}\label{eq:initL}
 \Tr_{\rm B}\argc{\wt H_{\rm int}(t),\r(0)}=\sum_{i,\m}\argc{ S_i^\m(t),\r_{\rm S}(0)} \text{Tr}\left( B_i^\m \rho_B\right)=0
\end{equation}

The second approximation, after the weak coupling, is a \textit{Markov approximation}:
\begin{equation}\label{eq:markov}
  \dot \r_{\rm S}(t)=-\int_0^\io\dd \t\, \Tr_{\rm B}\argc{\wt H_{\rm int}(t),[\wt H_{\rm int}(t-\t),\r_{\rm S}(t)\otimes \r_{\rm B}]}
\end{equation}
The latter approximation amounts to say that products like $\wt H_{\rm int}(t)\wt H_{\rm int}(t-\t)$ decay very rapidly to zero with $\t$ compared to the relaxation time $\t_{\rm R}$ of the system (the relaxation time of $ \r_{\rm S}(t)$), which is the case if the bath correlation functions (whose relaxation time is $\t_{\rm B}$) 
decay very fast: $\t_{\rm B}\ll\t_{\rm R}$. The evolution is now Markovian. 
Note that there is not a unique way to perform this approximation, as there is a wide freedom in the choice of the time substitution 
made above $t-\t\to t$ : for instance any time $t'$ such that $|(t-\t)-t'|\lesssim \t_{\rm B}$ could be chosen.

\subsection{Secular approximation and the Gorini-Kossakowski-Sudarshan-Lindblad equation}\label{sec:GKSL}

We shall now see that Eq.~\eqref{eq:markov} can be simplified further
\footnote{Let us note $\r_{\rm S}^s=\Tr_{\rm B}\r_{\rm tot}$ the Schr\"odinger picture density matrix. 
We have
\begin{equation}\label{eq:schrpic}
\begin{split}
 \r_{\rm S}(t)=&\Tr_{\rm B}\argp{U_0^\dagger(t)\r_{\rm tot}(t)U_0(t)}\\
 =&e^{it H_{\rm S}}\Tr_{\rm B}\argp{e^{it H_{\rm B}}\r_{\rm tot}(t)e^{-it H_{\rm B}}} e^{-it H_{\rm S}}\\
 =&e^{it H_{\rm S}}\r_{\rm S}^s(t)e^{-it H_{\rm S}}
\end{split}
 \end{equation}
\ie the standard relationship between Heisenberg and Schr\"odinger pictures for the reduced system S. 
If we switch to the Schr\"odinger picture, Eq.~\eqref{eq:markov} becomes
\clearpage
\begin{eqnarray}
 \dot \r_{\rm S}^s(t)=&&-i\argc{H_{\rm S},\r_{\rm S}^s(t)}\nonumber\\\nonumber
 +\sum_{\substack{i,j\\\m,\n}}\int_0^\io&&\dd \t\argp{ S_j^\n(-\t)\r_{\rm S}^s(t) S_i^\m- S_i^\m  S_j^\n(-\t)\r_{\rm S}^s(t)}\moy{ B_i^\m(\t) B_j^\n}_{\rm B} \\
 &&+ \ \textrm{H.c.}
\end{eqnarray}
We used time-translation invariance of the equilibrium bath correlation function. One sees that the extra difficulty of evolving in time the interaction Hamiltonian remains even in this formulation, which calls for the eigendecomposition of $H_{\rm S}$.} using the eigendecomposition of $H_{\rm S}$. 
We define the projector $\Pi(\varepsilon)$ on the eigenspace associated to the energy $\varepsilon$ of $H_{\rm S}$, and for each gap $\om$ of the spin Hamiltonian
\begin{equation}\label{eq:defprojom}
  S_i^\m(\om)\equiv\sum_{\varepsilon'-\varepsilon=\om} \Pi(\varepsilon) S_i^\m\Pi(\varepsilon') \qquad \Rightarrow\qquad H_{\rm int}=\sum_{i,\m,\om}  S_i^\m(\om)\otimes  B_i^\m
\end{equation}
This allows to simply write
\begin{equation}
 \wt H_{\rm int}(t)=\sum_{i,\m,\om}  e^{-i\om t} S_i^\m(\om)\otimes B_i^\m(t)=\sum_{i,\m,\om}  e^{i\om t} S_i^\m(\om)^\dagger\otimes  B_i^{\m\dagger}(t)
\end{equation}
with $ S_i^\m(\om)^\dagger= S_i^\m(-\om)$ from Eq.~\eqref{eq:defprojom}. Plugging it in Eq.~\eqref{eq:markov} using the Hermitian conjugated form for $\wt H_{\rm int}(t)$ and the direct one for $\wt H_{\rm int}(t-\t)$, we get
\begin{equation}\label{eq:markovrhos}
 \dot \r_{\rm S}(t)=\sum_{\substack{i,j\\\m,\n}}\sum_{\om,\om'}\G_{ij}^{\m\n}(\om)e^{i(\om'-\om)t}\argp{ S_j^\n(\om)\r_{\rm S}(t) S_i^\m(\om')^\dagger- S_i^\m(\om')^\dagger  S_j^\n(\om)\r_{\rm S}(t)}+\text{H.c.} 
\end{equation}
The equilibrium bath correlation functions $\G_{ij}^{\m\n}(\om)$ satisfy time-translational invariance owing to the commutation $\argc{H_{\rm B},\r_{\rm B}}=0$, so that using 
$\moy{B_i^\m(t)B_j^\n(t-\t)}=\moy{B_i^\m(\t)B_j^\n(0)}$, they are defined as
\begin{equation}\label{eq:corrfbath}
 \G_{ij}^{\m\n}(\om)=\int_0^\io\dd \t\,e^{i\om\t}\moy{B_i^\m(\t)B_j^\n(0)}_{\rm B}=\d_{ij}\d_{\m\n}\G(\om)
\end{equation}
We assumed as previously mentioned that different degrees of freedom of the bath decouple. 
Let us then write Eq.(\ref{eq:markovrhos}) back to the Schr\"odinger picture (see Eq.~\eqref{eq:schrpic}):
\begin{equation}\label{eq:markovrhos2}
\begin{split}
 \dot \r_{\rm S}^s(t)=-i\argc{H_{\rm S},\r_{\rm S}^s(t)}+\sum_{i,\m}\sum_{\om,\om'}&\left\{\G(\om)\argp{S_i^\m(\om)\r_{\rm S}^s(t)S_i^\m(\om')^\dagger-S_i^\m(\om')^\dagger S_i^\m(\om)\r_{\rm S}^s(t)} +\text{H.c.} \right\}\\
 \end{split}
\end{equation}
where we note that the phases disappear compared to the Heisenberg picture equation (\ref{eq:markovrhos}).
Eq.(\ref{eq:markovrhos2}) is written using an eigenbasis of $H_{\rm S}$. 
This is a convenient choice, but we could have chosen a different basis, yielding a more complicated equation, and at this level of approximation the time evolution of $\rho_{\rm S}^s$ would be identical. 

Nonetheless our Markovian quantum master equation must preserve the defining properties of a density matrix (such as positivity and trace), \ie be of the Gorini-Kossakowski-Sudarshan-Lindblad
(GKSL) form $\dot \r=\LL\r$ where the superoperator $\LL$ is a generator of a quantum dynamical semi-group~\cite[Sec.3.2]{BP02}. A more explicit definition is given by Eq. (4) of the main text. 
This is not guaranteed by Eq.~\eqref{eq:markovrhos2}. To ensure so in this weak-coupling approach, the standard prediction resorts to the \textit{secular approximation}. It amounts to
neglect the terms for which $\om\neq\om'$. It is then useful to define the real and imaginary parts:
\begin{equation}\label{eq:defReIm}
\begin{split}
  \G(\om)=&\frac{\g(\om)}{2}+i S(\om)\ ,\quad S(\om)= \Im\G(\om)\\
   \g(\om)=& 2\Re\G(\om)=\int_{-\io}^\io\dd \t\,e^{i\om\t}\moy{B_i^\m(\t)B_i^\m(0)}_{\rm B}
\end{split}
\end{equation}
As $\r_{\rm B}$ describes the Boltzmann-Gibbs distribution at inverse temperature $\b$, one can easily prove 
the Kubo-Martin-Schwinger condition $\moy{B_i^{\m\dagger}(t)B_j^\n(0)}_{\rm B}=\moy{B_j^\n(0)B_i^{\m\dagger}(t+i\b)}_{\rm B}$, from which the detailed balance condition 
$\g(\om)/\g(-\om)=e^{\b\om}$ follows~\cite[Sec.12]{pottier}. We can thus equivalently define
\begin{equation}\label{eq:detbal}
 \g(\om)=\frac{h(\om)}{T(|\om|)} \qquad\text{with}\qquad h(\om)=\frac{e^{\b\om}}{1+e^{\b\om}}
\end{equation}
$T(|\om|)$ defines relaxation times of the system for transitions at energy $\om$ while $h$ enforces the detailed balance condition.
Inserting the above definitions in Eq.~\eqref{eq:markovrhos2} where the $\om\neq\om'$ terms are neglected, one arrives at a GKSL master equation:
\begin{equation}\label{eq:LindbladS}
\begin{split}
    \dot \r_{\rm S}^s(t)=&-i\argc{H_{\rm S}+H_{\rm LS},\r_{\rm S}^s(t)}+\DD \r_{\rm S}^s(t)\\
   H_{\rm LS}=&\sum_{i,\m,\om}S(\om)S_i^{\m\dagger}(\om)S_i^\m(\om)\\
   \DD \r=&\sum_{i,\m,\om}\frac{h(\om)}{T(|\om|)}\argp{S_i^\m(\om)\r S_i^{\m\dagger}(\om)-\frac12\arga{S_i^{\m\dagger}(\om)S_i^\m(\om),\r}}
\end{split}
\end{equation}
The \textit{Lamb-shift} Hamiltonian $H_{\rm LS}$ being Hermitian and $\argc{H_{\rm S},H_{\rm LS}}=0$,  it can be absorbed into the system Hamiltonian $H_{\rm S}$ in Eq.~\eqref{eq:LindbladS} just by shifting the energy levels. 

For our many-body electron system described by the Hamiltonian $H_{\rm S}$, it is safe to consider 
the energy gaps as non degenerate. Eq.~\eqref{eq:LindbladS} takes then a simple form in the eigenbasis $\{\ket{n}\}$ 
\begin{equation}\label{eq:unitarybath}
 \begin{split}
  \dot \r_{nn}=&\sum_{m\neq n}W_{mn}^{\rm B}\r_{mm}-W_{nm}^{\rm B}\r_{nn}\\
  \dot \r_{m\neq n}=&-\argp{i\om_{mn}+\frac{1}{T_{mn}}}\r_{m\neq n}
 \end{split}
\end{equation}
with respectively the bath and decoherence rates
\begin{equation}\label{eq:bathrate}
 \begin{split}
 W_{nm}^{\rm B}=&\frac{h(\om_{nm})}{T(|\om_{nm}|)}\sum_{i,\m}\abs{\bra{n}S_i^\m\ket{m}}^2\\
  \frac{1}{T_{nm}}=& \frac12\sum_{k\neq n}h(\om_{nk})W_{nk}^{\rm B}+\frac12\sum_{k\neq m}h(\om_{mk})W_{mk}^{\rm B}
  +\frac{1}{4T(0)}\sum_i \bra{n}S_i^z\ket{n}^2+\bra{m}S_i^z\ket{m}^2-2\bra{n}S_i^z\ket{n}\bra{m}S_i^z\ket{m}
 \end{split}
\end{equation}
 The first line in Eq.~\eqref{eq:unitarybath} is a master equation for the diagonal elements only, 
with rates satisfying the detailed balance condition $W^{\rm B}_{mn}/W^{\rm B}_{nm}=e^{\b\om_{mn}}$, and 
$\om_{mn}=\varepsilon_m-\varepsilon_n$ the gap between levels $m$ and $n$ of the Hamiltonian $H_{\rm S}+H_{\rm LS}$. 
The second equation describes the evolution of the off-diagonal terms: the first term $i\om_{mn}$ represents the dephasing from the unitary evolution 
while the second term $T_{mn}^{-1}$ describes the decoherence due to the bath.
Under this evolution, the coherences vanish exponentially while the populations tend to their thermal values imposed by the Boltzmann distribution~\cite[Sec.3.3.2]{BP02}, emerging from the detailed balance at temperature $\b^{-1}$.

Note that \textit{after} the secular approximation, the time evolution of $\r_{\rm S}^s$ depends on the choice of the eigenbasis. 
In presence of a drive, different steady states can be reached by changing this choice of basis, which is an inconsistency of the secular approximation.

\subsection{The nonsecular quantum master equation}\label{sub:NR20}

A usual justification~\cite{BP02} for the secular approximation, on top of allowing to preserve the properties of a density matrix, is that in Heisenberg representation the terms $\om\neq\om'$ produce rapidly oscillating phases (see Eq.~\eqref{eq:markovrhos}) which cancel (note however that such phases are absent in Schr\"odinger representation). Yet for a large many-body system there exist a huge number of quasi-degenerate gaps for which the argument does not hold. 
There is a way to retain all the terms from Eq.~\eqref{eq:markovrhos} while ensuring a GKSL form by modifying the Markovian approximation and not having to perform any additional approximation. Below we summarize the derivation as given by Nathan and Rudner's recent paper~\cite{NR20}.

\subsubsection{Derivation of the nonsecular equation}\label{sub:NRrev}

We start back from Eqs.~\eqref{eq:quasiRedfield}-\eqref{eq:markov}. Coming back to the variable $s$ we see that the initial condition is in the end preponed to $-\io$ by the Markov approximation performed there, and we have
\begin{equation}\label{eq:BMNR}
  \dot\r_{\rm S}(t)=-\int_{-\io}^t\dd s\, \Tr_{\rm B}\argc{\wt H_{\rm int}(t),[\wt H_{\rm int}(s),\r_{\rm S}(t)\otimes\r_{\rm B}]}
  =-\sum_{i,\m} \int_{-\io}^t\dd t'\,\G(t-t')\argc{S_i^\m(t),S_i^\m(t')\r_{\rm S}(t)}\ +\ \textrm{H.c.}
\end{equation}
A first progress is to take the square root of the bath rate in Fourier space, \ie to define $g$ such that 
\begin{equation}\label{eq:defsqrt}
 \G(t-t')=\int_\RRR\dd s\, g(t-s)g(s-t')
\end{equation}
Eq.~\eqref{eq:BMNR} becomes
\begin{equation}
\begin{split}
  \dot\r_{\rm S}(t)=&\int_{\RRR^2}\dd t'\dd s\, \FF(t,s,t')[\r_{\rm S}(t)]\\
   \FF(t,s,t')[\r]=&\sum_{i,\m}\th(t-t')g(t-s)g(s-t')\argc{S_i^\m(t),S_i^\m(t')\r}\ +\ \textrm{H.c.}
 \end{split} 
\end{equation}
Integrating on $[t_1,t_2]$ with $t_2-t_1\gg \t_{\rm B}$:
\begin{equation}\label{eq:NR1}
 \r_{\rm S}(t_2)-\r_{\rm S}(t_1)=\int_{t_1}^{t_2}\dd t\int_{\RRR^2}\dd t'\dd s\,\FF(t,s,t')[\r_{\rm S}(t)]
\end{equation}
Next, due to $t_2-t_1\gg \t_{\rm B}$, most contributions in the integrals of Eq.~\eqref{eq:NR1} arise for $(t,s,t')\in[t_1,t_2]^3$, thus they are approximately unaffected if we change the boundaries as
$\{-\io<s,t'<\io,t_1\leqslant t \leqslant t_2\}\to\{-\io<t,t'<\io,t_1\leqslant s \leqslant t_2\}$:
\begin{equation}\label{eq:NR2}
 \r_{\rm S}(t_2)-\r_{\rm S}(t_1)\simeq\int_{t_1}^{t_2}\dd s\int_{\RRR^2}\dd t\dd t'\,\FF(t,s,t')[\r_{\rm S}(s)]
\end{equation}
where, similarly to Eq.~\eqref{eq:markov} one makes another Markovian type of approximation for the density matrix time dependence. Then we derive Eq.~\eqref{eq:NR2} with respect to $t_2$ and set $t=t_2$:
\begin{equation}
 \dot\r_{\rm S}(t)=\LL(t)[\r_{\rm S}(t)]\ ,\qquad 
 \LL(t) = \int_{\RRR^2}\dd s\dd s'\,\FF(s,t,s')
\end{equation}
and using $\th(t)=1/2 + \sign(t)/2$ helps us to disentangle the Lamb-shift and the jump operators:
\begin{equation}
 \begin{split}
  \dot\r_{\rm S}(t)=&-i\argc{H_{\rm LS}(t),\r_{\rm S}(t)}+\sum_{i,\m} A_i^\m(t)\r_{\rm S}(t)A_i^\m(t)^\dagger
  -\frac12\arga{A_i^\m(t)^\dagger A_i^\m(t),\r_{\rm S}(t)}\\
  A_i^\m(t)=&\int_\RRR\dd s\, g(t-s)S_i^\m(s)\ ,\qquad 
  H_{\rm LS}(t)=\frac{1}{2i}\sum_{i,\m} \int_{\RRR^2}\dd s\dd s'\,
  S_i^\m(s)g(t-s)g(t-s')S_i^\m(s')\sign(s-s')
 \end{split}
\end{equation}
By Fourier transform and the definition~\eqref{eq:defsqrt} we can relate $g(t)$ and $\g(\om)$:
\begin{equation}
\tilde g(\om)=\int_\RRR\frac{\dd t}{2\p}\,e^{i\om t}g(t) \quad \Rightarrow \quad \tilde g(\om)=\frac{\sqrt{\g(\om)}}{2\p}
\end{equation}
We now seek to come back to Schr\"odinger representation and decompose the operators on the eigenbasis of the spin Hamiltonian through $S_i^\m(t)=\sum_\om S_i^\m(\om)e^{-i\om t}$. For the Lamb-shift Hamiltonian
we show through the Fourier decompositions
\begin{equation}
\begin{split}
  H_{\rm LS}(t)=&\sum_{i,\m,\om,\om'}S_i^\m(\om)S_i^\m(\om')e^{-it(\om+\om')}
 \int_{\RRR^2}\dd \Omega\dd \Omega'\,\tilde g(\Omega)\tilde g(\Omega')k(\om+\Omega,\om'-\Omega')\\
 k(p,q)=&\frac{1}{2i}\int_{\RRR^2}\dd s\dd s'\,\sign(s-s')e^{-i(ps+qs')}
\end{split}
 \end{equation}
Through a direct computation and usual regularizations we get
\begin{equation}
 k(p,q)=2\p\d(p+q)\Im\frac{1}{ip+0^+}
\end{equation}
which yields
\begin{equation}
  H_{\rm LS}(t)=\sum_{i,\m,\om,\om'}S_i^\m(\om)S_i^\m(\om')e^{-it(\om+\om')}
2\p\PP\int_\mathbbm{R} \dd\Omega\,\frac{\tilde g(\Omega-\om)\tilde g(\Omega+\om')}{\Omega}
\end{equation}
where $\PP$ stands for Cauchy's principal value. 
It is then straightforward to pass from Heisenberg to Schr\"odinger representations, which cancels the phases. This allows to define nonsecular jump operators 
\begin{equation}\label{eq:nonsecjumpop}
 A_i^\m=\sum_\om \sqrt{\g(\om)}S_i^\m(\om)
\end{equation}
In Schr\"odinger representation, the nonsecular version of Eq.~\eqref{eq:LindbladS} thus reads:
\begin{equation}\label{eq:NR20}
 \begin{split}
   \dot \r_{\rm S}^s(t)=&-i\argc{H_{\rm S}+H_{\rm LS},\r_{\rm S}^s(t)}+
   \sum_{i,\m}A_i^\m\r_{\rm S}^s(t)A_i^{\m\dagger}
   - \frac12\arga{A_i^{\m\dagger}A_i^\m ,\r_{\rm S}^s(t)}\\
   H_{\rm LS}=&\sum_{i,\m,\om,\om'}S_i^\m(\om)S_i^\m(\om')f(\om,\om')\ ,\qquad
  f(\om,\om') =\PP\int_\mathbbm{R} \frac{\dd\Omega}{2\p}\,\frac{\sqrt{\g(\Omega-\om)\g(\Omega+\om')}}{\Omega}
 \end{split}
\end{equation}
The Lamb-shift $H_{\rm LS}$ is Hermitian and the evolution is thus manifestly in GKSL form. The secular equation is recovered again by throwing away all gaps $\om\neq\om'$ in the sums. 

\subsubsection{Lamb shift and nonsecular jump operators}\label{sub:NRLS}
For our system the Lamb-shift $H_{\rm LS}$ becomes negligible due to the wide difference of characteristic energy scales. Indeed considering that $f(\om,\om')$ is smooth in both arguments, $f(\om,\om')\simeq 
f(0,0)$ for $S_i^z$ gaps while for $S_i^{x,y}$ we can use $S_i^\pm=S_i^x\pm i S_i^y$ operators and get:
\begin{equation}
 \begin{split}
  H_{\rm LS}\simeq& f(0,0) \sum_i\argp{S_i^z}^2 +\sum_{i,\om,\om'}\frac14\argp{S_i^+(\om)+S_i^-(\om)}\argp{S_i^+(\om')+S_i^-(\om')} 
 f(\om,\om')\\
 &-\sum_{i,\om,\om'}\frac14\argp{S_i^-(\om)-S_i^+(\om)}\argp{S_i^-(\om')-S_i^+(\om')}f(\om,\om') \\
 = & \frac N 4 f(0,0) +\frac12\sum_{i,\om,\om'}f(\om,\om')\argc{S_i^+(\om)S_i^-(\om')+S_i^-(\om)S_i^+(\om')}\\
 \simeq&\frac N 4 f(0,0) +\frac{1}{2}\sum_{i,\om,\om'}\argc{f(-\ome,\ome)S_i^+(\om)S_i^-(\om')+f(\ome,-\ome)S_i^-(\om)S_i^+(\om')}\\
 =&\frac N 4 f(0,0) +\frac{f(\ome,-\ome)}{2}\sum_i\argp{S_i^+S_i^-+S_i^-S_i^+}
 =\frac N 4(f(0,0)+2f(\ome,-\ome))\mathbbm{1}
 \end{split}
\end{equation}
where we have used the symmetry $f(\ome,-\ome)=f(-\ome,\ome)$. In conclusion the Lamb-shift Hamiltonian is proportional to the identity up to very small corrections and thus we can take 
$\argc{H_{\rm LS},\r_{\rm S}^s(t)}=0$ in Eq.~\eqref{eq:NR20}. We simply discarded Lamb-shifts in the main text.

With very similar considerations one obtains from the definition~\eqref{eq:nonsecjumpop} the expressions of the nonsecular jump operators given by Eqs. (7)-(8) of the main text.

\subsection{Turning the microwaves on}\label{sec:MW}

In the following we set $H_{\rm LS}=0$ and incorporate the microwave field\footnote{If the field is 
oscillating along a single direction, \eg  $H_{\rm MW}(t)=\om_1S^x \cos(\mw t)$, the conclusion of this section remains valid at small drive through the rotating wave approximation~\cite{DLR15,DLRAMR16}.}  
\begin{equation}
 H_{\rm MW}(t)=\om_1\argc{S^x \cos(\mw t)+S^y \sin(\mw t)}
\end{equation}
The interaction picture dynamics Eq.~\eqref{eq:vN} becomes
\begin{equation}
 \dot \r(t)=-i[\wt H_{\rm MW}(t),\r(t)]-i[\wt H_{\rm int}(t),\r(t)]
\end{equation}
with $\wt H_{\rm MW}(t)=U_0^\dagger(t) H_{\rm MW}(t)U_0(t)=e^{iH_{\rm S}t}H_{\rm MW}(t)e^{-iH_{\rm S}t}$. Then we treat the last term as before by integrating once the equation and obtain
\begin{equation}\label{eq:perturbMW}
 \dot \r(t)=-i[\wt H_{\rm MW}(t),\r(t)]-\int_0^t \dd s\,\argc{\wt H_{\rm int}(t),[\wt H_{\rm int}(s),\r(s)]}-\int_0^t \dd s\,\argc{\wt H_{\rm int}(t),[\wt H_{\rm MW}(s),\r(s)]}-i[\wt H_{\rm int}(t),\r(0)]
\end{equation}
Then if we make the Born approximation $\r(t)\simeq \r_{\rm S}(t)\otimes \r_{\rm B}$ and trace over the bath, since $\wt H_{\rm MW}(s)$ acts only on the spin degrees of freedom, the last two terms in Eq.~\eqref{eq:perturbMW} are zero if the bath degrees of freedom are of average zero, $\moy{B_i^\m(t)}_{\rm B}=0$. Then the quadratic term in $H_{\rm int}$ can be dealt with as in the previous sections. 

In Schr\"odinger picture we therefore get the equation, similar to Eq.~\eqref{eq:NR20}:
\begin{equation}\label{eq:LMW}
\begin{split}
  \dot \r_{\rm S}^s(t)=&-i\argc{H_{\rm MW}(t),\r_{\rm S}^s(t)}-i\argc{H_{\rm S},\r_{\rm S}^s(t)}+\DD \r_{\rm S}^s(t)\\
  \DD \r=&-i\argc{H_{\rm LS},\r}+
   \sum_{i,\m}A_i^\m\r A_i^{\m\dagger}
   - \frac12\arga{A_i^{\m\dagger}A_i^\m ,\r}
  \end{split}
\end{equation}
Note that $H_{\rm MW}(t)$ is time dependent ; a usual trick for NMR studies is to place ourselves in a rotated frame given by the Larmor precession. 
We shall now rotate accordingly the reference frame, which, if done at frequency $\mw$, makes the microwaves field effectively stationary, resulting in a time-independent dynamical semigroup generator for Eq.~\eqref{eq:LMW}. 
The rotation is implemented as
\begin{equation}\label{eq:rotation}
 U_{\rm MW}(t)=e^{-i\mw t S^z}\ ,\qquad \r_r(t)=U_{\rm MW}^\dagger(t)\r_{\rm S}^s(t)U_{\rm MW}(t)
\end{equation}
Since $\argc{S^z,H_{\rm S}}=0$, the unperturbed spin Hamiltonian is unchanged. We then need to express the rotated microwave 
Hamiltonian $U_{\rm MW}^\dagger(t)H_{\rm MW}(t)U_{\rm MW}(t)$. To achieve this one can write the microwave Hamiltonian as
\begin{equation}
 H_{\rm MW}(t)=\frac{\om_1}{2}\argp{S^- e^{i\mw t}+S^+ e^{-i\mw t}}
\end{equation}
and  we define  $S^\pm(t)=U_{\rm MW}^\dagger(t)S^\pm U_{\rm MW}(t)$ verifying 
\begin{equation}
   \dot S^\pm(t)=U_{\rm MW}^\dagger(t)i\mw\argc{S^z,S^\pm}U_{\rm MW}(t)=\pm i\mw S^\pm(t)\qquad
  \Rightarrow\qquad S^\pm(t)=e^{\pm i \mw t}S^\pm
\end{equation}
where we used $\argc{S^z,S^\pm}=\pm S^\pm$. 
We thus conclude that in the rotating frame the microwave Hamiltonian becomes stationary:
\begin{equation}
 U_{\rm MW}^\dagger(t)H_{\rm MW}(t)U_{\rm MW}(t)=\om_1S^x
\end{equation}
and the GKSL equation~\eqref{eq:LMW} becomes
\begin{equation}\label{eq:rotfraderiv}
 \dot \r_r(t)=-i\argc{H_{\rm S}+\om_1S^x-\mw S^z,\r_r(t)}+ U_{\rm MW}^\dagger(t)\DD \argp{ U_{\rm MW}(t)\r_r(t) U_{\rm MW}^\dagger(t)} U_{\rm MW}(t)
\end{equation}
 Let us define as before $H_{\rm S}\ket{n}=\varepsilon_n\ket{n}$ the eigendecomposition. 
As $[S^z,H_{\rm S}]=0$, the total polarization along $z$ is conserved, and we note $S^z\ket{n}=s_n^z\ket{n}$. We consider as well for large enough $N$ that the finite gaps 
$\omega_{nm}=\varepsilon_n-\varepsilon_m$ are non degenerate, implying that the jump operators read:
\begin{equation}\label{eq:jumpopdecomp}
 S_i^z(0)=\sum_n\bra{n}S_i^z\ket{n}\ketbra{n}{n} \ , \qquad S_i^\m(\omega_{nm})=\ketbra{m}{m}S_i^\m\ketbra{n}{n}
\end{equation}
To analyze the last term of Eq.~\eqref{eq:rotfraderiv}, let us go back to the original dissipative part of Eq.~\eqref{eq:LMW} (second line). It consists in sums over pairs of gaps $(\om,\om')$ of the jumps operators~\eqref{eq:jumpopdecomp}. Each term in this sum contains a product of the form
\begin{equation}\label{eq:tr}
 \bra{n}S^\m_i(\om=\om_{mn})\ket{m}\bra{m'}S^\m_i(\om'=\om_{m'n'})^\dagger\ket{n'}
\end{equation}
The bath operators are of two kinds: $S^x_i$, $S^y_i$ which induce flips of the spin $i$ with energy jump $\approx\pm\om_{\rm e}$ 
and $S^z_i$ whose transitions have gaps $\om\ll\ome$ which conserve the total polarization along $z$.
For the $S^x_i$, $S^y_i$ operators, there are 4 possibilities:
\begin{itemize}
 \item $\om\approx \om_{\rm e}$ and $\om'\approx -\om_{\rm e}$: 
 both factors imply a flip $\ket{-}_i \to \ket{+}_i$, so that  ${i\bra{n}S^x_i\ket{m}=\bra{n}S^y_i\ket{m}}$ and ${i\bra{m'}S^x_i\ket{n'}=\bra{m'}S^y_i\ket{n'}}$. Consequently the terms generated by the operator $S_i^y$ have the opposite value to the one generated by $S_i^x$: 
 and their contribution vanishes.
 \item $\om\approx -\om_{\rm e}$ and $\om'\approx \om_{\rm e}$: 
 both factors imply a flip $\ket{+}_i \to \ket{-}_i$, so that  ${-i\bra{n}S^x_i\ket{m}=\bra{n}S^y_i\ket{m}}$ and ${-i\bra{m'}S^x_i\ket{n'}=\bra{m'}S^y_i\ket{n'}}$.
Once again their combined contribution vanishes.
 \item $\om\approx\om'(\approx\pm\om_{\rm e})$: contrary to the above cases those terms are allowed. 
\end{itemize}
For all allowed transitions the total polarization jump for the term~\eqref{eq:tr} 
(\ie from $\ket n$ to $\ket n'$) is $s_n^z-s_m^z-(s_{n'}^z-s_{m'}^z)=0$. 
This holds for the $S^x_i$, $S^y_i$ operators (for which $s_n^z-s_m^z=s_{n'}^z-s_{m'}^z=\pm1$), 
but also for $S^z_i$ transitions where $s_n^z=s_m^z$ and $s_{n'}^z=s_{m'}^z$. 

When shifting to the rotating frame, the rotation~\eqref{eq:rotation} produces oscillating phases in the dissipative term of Eq.~\eqref{eq:rotfraderiv} according to the following relation
\begin{equation}\label{eq:mwspin}
 U_{\rm MW}(t)S_i^\m(\om_{mn})U_{\rm MW}^\dagger(t)=e^{i\mw t(s^z_n-s^z_m)}S_i^\m(\om_{mn})
\end{equation}
Therefore each term writing as~Eq.\eqref{eq:tr} receives a phase 
$e^{i\mw t\argc{s^z_n-s^z_m-(s^z_{n'}-s^z_{m'})}}$. As emphasized above, this phase is 1 for all allowed transitions. 
Consequently $U_{\rm MW}^\dagger(t)\DD \argp{ U_{\rm MW}(t)\r_r(t) U_{\rm MW}^\dagger(t)} U_{\rm MW}(t)=\DD\r_r(t)$, \ie the quantum master equation in the rotating frame is
\begin{equation}\label{eq:Lindrot}
 \dot \r_r(t)=-i\argc{H_{\rm S}+H_{\rm LS}+\om_1S^x-\mw S^z,\r_r(t)}
 + \sum_{i,\m}A_i^\m\r_r(t)A_i^{\m\dagger}
   - \frac12\arga{A_i^{\m\dagger}A_i^\m ,\r_r(t)}
\end{equation}
Note that if for example only the spin $j$ is irradiated, one has to replace $S^x\to S^x_j$.

\subsection{The Hilbert approximation}\label{sub:Hilb}

\begin{figure}[t]
 \includegraphics[width=0.6\linewidth]{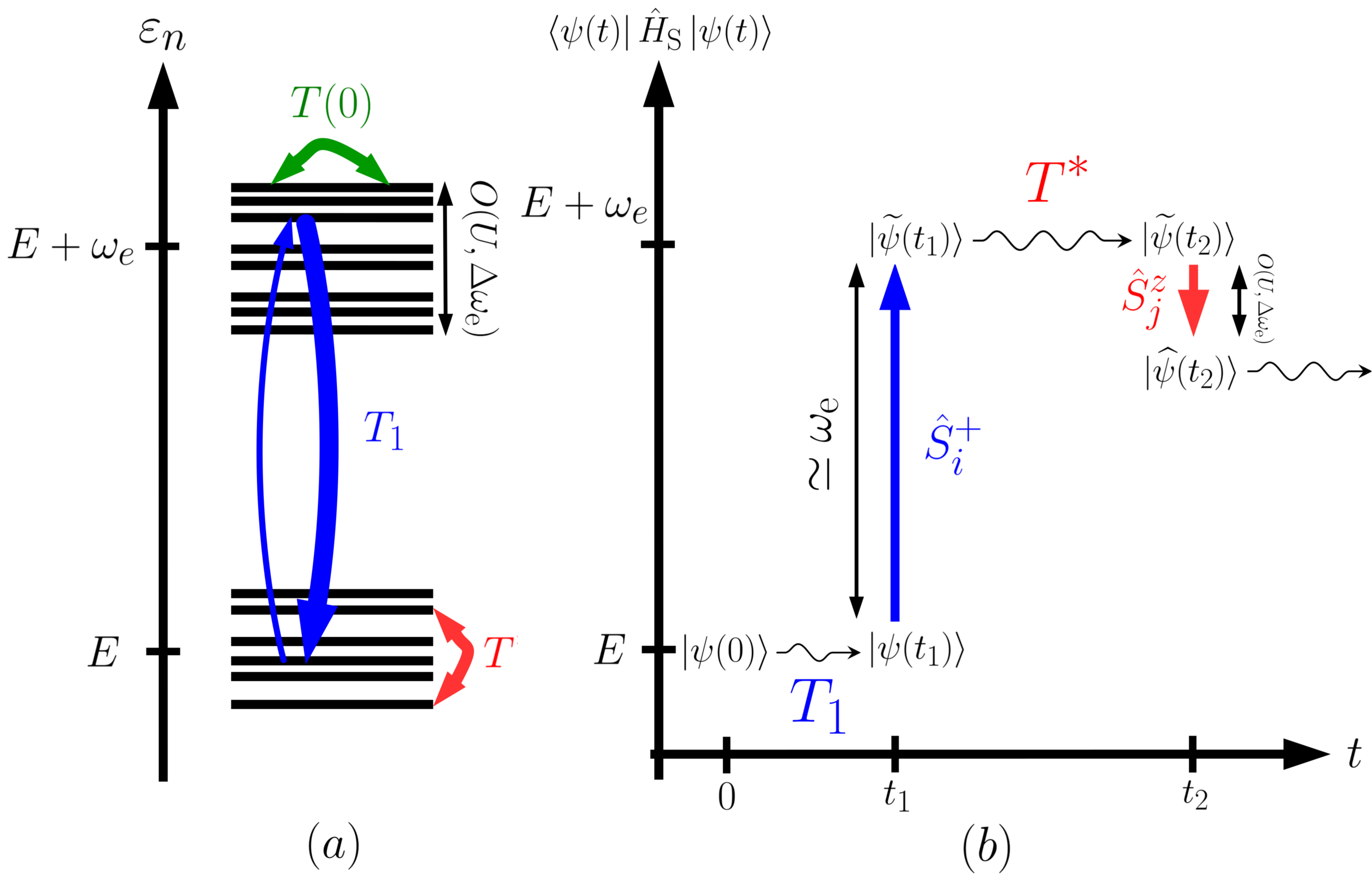}
 \caption{Effective quantum dynamics of the electron spins. 
  (a) Energy levels of $  H_{\rm S}$ are drawn. They are arranged into $N+1$ polarization sectors, each containing  many levels, separated by multiples of $\ome$. 
  In the Hilbert approximation due to fast dephasing, the jump operators (arrows)  select transitions between eigenstates with corresponding timescales, and the system is a mixture of $  H_{\rm S}$ eigenstates.
  (b) Example of a nonsecular quantum trajectory in its average energy vs time plane: the unitary dynamics (wavy arrows) gets projected with a characteristic rate by  spatially-localized jump operators. The Hilbert approximation is retrieved for slow jump rates for which the unitary evolution is able to project the system onto eigenstates.}
 \label{fig:dyn}
\end{figure}

In this section we provide the details of the Hilbert approximation for the secular GKSL equation 
given by Eq.~\eqref{eq:Lindrot} where nonsecular terms $\om\neq\om'$ are thrown away (as in Eq.~\eqref{eq:LindbladS}), which amounts to take $A_i^\m\to \sqrt{\g(\om)}S_i^\m(\om)$ in Eq.~\eqref{eq:Lindrot}. Below we drop the rotating frame index $r$ and consider $H_{\rm LS}=0$ as explained in the previous sections:
\begin{equation}\label{eq:secularL}
 \dot \r=-i[H_{\rm S}-\mw S^z+\om_1S^x,\r]+\sum_{\om,i,\m}\frac{h(\om)}{T(|\om|)}\argp{S_i^\m(\omega)\r S_i^\m(\omega)^\dagger-\frac12\arga{S_i^\m(\omega)^\dagger S_i^\m(\omega),\r}}
\end{equation}
Projecting on the diagonal and off-diagonal terms ($n\neq m$) leads to
\begin{equation}\label{eq:secproj}
\begin{split}
  \dot\r_{nn}=&\sum_{k\neq n}W_{kn}^{\rm B}\r_{kk}-W_{nk}^{\rm B}\r_{nn}
  -i\om_1\argp{\bra{n}S^x\ket{k}\r_{kn} -\ \textrm{c.c.}}\\
  \dot\r_{nm}=&  -\argp{i\D\om_{nm}+\frac{1}{T_{nm}}}\r_{nm}
  -i\om_1\sum_{k}\argp{\bra{n}S^x\ket{k}\r_{km} -\bra{k}S^x\ket{m}\r_{nk} }
\end{split}
\end{equation}
where rates have been defined in Eq.~\eqref{eq:bathrate} and the dephasing is now 
\begin{equation}
 \D\om_{nm}=\varepsilon_n-\varepsilon_{m}-(s^z_n-s^z_{m})\mw 
\end{equation}
If the term $i\D\om_{nm}+1/T_{nm}$ dominates the coherences' evolution -- making them vanish exponentially fast -- the dynamics is projected on the diagonal elements. This is the spirit of the \textit{Hilbert approximation}. It is achieved by considering $T_{nm}$ and $1/\D\om_{nm}$ go to zero with  $\D\om_{nm}T_{nm}$ constant and solving perturbatively the master equation. We thus write the evolution as $\dot\r=L_0\r+L_1\r$ with $L_0$ the dominant contribution to the generator, defined by
\begin{equation}
 L_0 e_{nm}=\left\{
 \begin{split}
  &0 \qquad n=m\\
  &-\argp{i\D\om_{nm}+\frac{1}{T_{nm}}}e_{nm} \qquad n\neq m
 \end{split}
 \right.
\end{equation}
with $\{e_{nm}\}$ a basis for our $2^N\times 2^N$ density matrices: $\argp{e_{nm}}_{ab}=\d_{an}\d_{bm}$. 
At dominant order the steady-state density matrices fall in the subspace of diagonal matrices. Treating $L_1$ as a perturbation, the Hilbert approximation projects the dynamical evolution in the diagonal subspace, leading to an approximate master equation for the populations. Details of the (Schrieffer-Wolff) perturbation theory are described in~\cite{DLR15,RARDL18,coleman}.
One has to go to second order in $L_1$ to get the first terms dependent on $\D\om_{nm}$ and $T_{nm}$.
One finds  the master equation for the populations
\begin{equation}\label{eq:masterpn}
 \dot \r_{nn}=\sum_{n'}\argc{(L_1)_{nn,n'n'}+ \sum_{m\neq m'}\frac{(L_1)_{nn,mm'}(L_1)_{mm',n'n'}}{i\D\om_{mm'}+1/T_{mm'}}}\r_{n'n'}
\end{equation}
with the basis decomposition $(L_1\r)_{nm}=\sum_{ij}(L_1)_{nm,ij}\r_{ij}$. 
Concretely the Hilbert rate equation~\eqref{eq:masterpn} reads in the present secular case
\begin{equation}\label{eq:hilbsec}
 \dot \r_{nn}=\sum_{m\neq n}\argc{W_{mn}^{\rm B}+W_{nm}^{\rm MW}}\r_{mm}-\argc{W_{nm}^{\rm B}+W_{nm}^{\rm MW}}\r_{nn}
\end{equation}
where the microwave rates are
\begin{equation}\label{eq:WMW}
 W_{nm}^{\rm MW}=\frac{2\om_1^2T_{nm}}{1+\argp{T_{nm}\D\om_{nm}}^2}\abs{\bra{n}S^x\ket{m}}^2
\end{equation}
The physical interpretation of Hilbert versus nonsecular dynamics is sketched in Fig.~\ref{fig:dyn}.

\subsection{Comparison between secular and nonsecular GKSL equations}\label{sub:comparison}

In this section we highlight the additional terms contained in the nonsecular GKSL equation~\eqref{eq:Lindrot} compared to the 
secular one~\eqref{eq:secularL} (with $H_{\rm LS}=0$). 
We write the off-diagonal elements of the nonsecular equation in the eigenbasis, 
as these contain the main new terms to account for the bath-induced localization. 
The first line corresponds to the secular part given in Eq.~\eqref{eq:secproj}:
\begin{equation}\label{eq:nonsecproj}
\begin{split}
  \dot\r_{nm}=&  -\argp{i\D\om_{nm}+\frac{1}{T_{nm}}}\r_{nm}
  -i\om_1\sum_{k}\argp{\bra{n}S^x\ket{k}\r_{km} -\bra{k}S^x\ket{m}\r_{nk} }\\
  &+\sum_{\substack{i\\\m=x,y,z}}\left[\sum_{\substack{k\neq n\\k'\neq m}}\sqrt{\g(\om_{kn})\g(\om_{k'm})}\underbrace{\bra{n}S^\m_i\ket{k}}_{\om=\om_{kn}}\r_{kk'}
  \underbrace{\bra{k'}S^\m_i\ket{m}}_{\om'=\om_{k'm}}
  -\frac12\sum_{\substack{k\\k'\neq n}}\sqrt{\g(\om_{nk})\g(\om_{k'k})}\underbrace{\bra{n}S^\m_i\ketbra{k}{k} S^\m_i\ket{k'}
  \r_{k'm}}_{\om'=\om_{nk},\ \om=\om_{k'k}}\right.\\
  &\left. \qquad\qquad\qquad-\frac12\sum_{\substack{k\\k'\neq m}}\sqrt{\g(\om_{k'k})\g(\om_{mk})} \r_{nk'}
  \underbrace{\bra{k'}S^\m_i\ketbra{k}{k} S^\m_i\ket{m}}_{\om'=\om_{k'k},\ \om=\om_{mk}}\right]
\end{split}
\end{equation}
Some remarks:
\begin{enumerate}
 \item The constraints over eigenstates indices below the sum signs correspond to secular terms already taken into account in the first line of the equation. 
 For instance if the constraint is $k\neq n$ this means that the same term with $k=n$ is a secular term contained in the first line.
 \item For $S_i^{x,y}$ terms, we have explicitly written the pair of gaps $(\om,\om')$ involved. This emphasizes, as remarked 
 in Sec.~\ref{sec:MW}, that terms not satisfying $\om\approx\om'$ actually vanish, although we have not mentioned it in the sums to lighten the notation. This is not true for  $S_i^z$ terms.
 \item The secular equation at small drive allows only a steady-state solution where all coherences $\r_{nm}=0$. 
 The nonsecular terms render possible non-zero coherences in the stationary state: the system does not get projected anymore onto 
 the Hamiltonian eigenstates. Considering the reasonable assumption $T^* < T_1$, this can occur if the timescale $T^*$ gets lowered so that the nonsecular terms in Eq.~\eqref{eq:nonsecproj} 
 do have an impact. This timescale comes from $S_i^z$ transitions. All nonsecular $S_i^z$ transitions in  Eq.~\eqref{eq:nonsecproj} occur 
 between different eigenstates ; as a consequence, in the limit of strong disorder or non-interacting spins, such terms vanish. 
 This explains why the drastic effect of the bath seen for an ETH Hamiltonian is not present in the case of a MBL Hamiltonian, see Fig~\ref{fig:MBL}: there, Hilbert dynamics is sufficiently accurate. 
 We stress that, formally considering $T_1 < T^*$ instead, a shortening of $T_1$ associated to $S_i^{x,y}$ transitions leads to a similar mechanism. This is shown in Fig.~\ref{fig:T1small}.
 
\end{enumerate}

\section{Electron Paramagnetic Resonance spectrum}\label{sec:EPR}
The aim of this section is to show how we compute numerically the EPR spectra. 
For this we need to extend the results of~\cite{CFCRDL16} to the nonsecular case in which the  
stationary density matrix $\r_{\rm stat}$ possesses non-zero coherences in the eigenbasis of the system Hamiltonian. 
This provides the formula~\eqref{eq:tildefexpr} for a smoothed EPR spectrum which is appropriate for numerical evaluation.

We recall that after the $\p/2$ pulse (performed in the stationary state at a time that we note $\t=0$) the density matrix gets transformed into 
$\r_{\p/2}=e^{i\frac\p 2 S_i^x}\r_{\rm stat}e^{-i\frac\p 2 S_i^x}$. 
The polarization of spin $i$ is given at later times by
\begin{equation}
 P_{i}^\m(\t)=2\Tr(S_i^\m(\t)\r_{\p/2})\ ,\qquad S_i^\m(\t)=e^{iH_{\rm S}\t}S_i^\m e^{-iH_{\rm S}\t}
\end{equation}
and we recall the quantity defined in Eq. (11) of the main text:
\begin{equation}\label{eq:polim}
 g_i(\t)=(P_i^y-iP_i^x)(\t)=-2i\Tr\argc{ S_i^+(\t)\r_{\p/2}}
\end{equation}
Note that in our formalism the steady-state density matrix $\r$ that we get numerically is expressed in the $(x,y)$ plane in the rotating frame at frequency $\mw$, whereas $\r_{\rm stat}$ is in the fixed frame.
Their relationship is ${\r_{\rm stat}=U_{\rm MW}(\t)\r U_{\rm MW}(\t)^\dagger}$. 

The polarization on the $z$ axis (which is an invariant axis) in the steady-state can be computed with either density matrix. Indeed,
\begin{equation}
 \Tr\argp{S_i^z\r_{\rm stat}}=\sum_{n,m} e^{i\mw(s_m^z-s_n^z)\t}\r_{nm}\bra{m}S_i^z\ket{n}
 =\sum_{\substack{n,m\\s_n^z=s_m^z}}\r_{nm}\bra{m}S_i^z\ket{n}=\Tr\argp{S_i^z\r}
\end{equation}
as $\bra{m}S_i^z\ket{n}$ vanishes outside the blocks of constant $S^z$ (defined by $s_m^z=s_n^z$).

In the following we focus on the EPR spectrum instead. First, to express the post-pulse polarization with $\r$ instead of $\r_{\p/2}$, we write
\begin{equation}
 e^{i\frac{\p}{2}S_i^x}=\cos\frac\p4 +2iS_i^x\sin\frac\p4=\frac{1}{\sqrt 2}\argp{\mathbbm{1}+2iS_i^x}
\end{equation}
Starting from the definition Eq.~\eqref{eq:polim} :
\begin{equation}
 g_i(\t)=-i\sum_{k,l,m,n}e^{i\t[\varepsilon_k-\varepsilon_l+\mw(s^z_m-s^z_n)]}
 \bra{m}1-2iS_i^x\ketbra{k}{k}S_i^+\ketbra{l}{l}1+2iS_i^x\ket{n}\r_{nm}
\end{equation}
We thus expect the frequency spectrum to be a sum of exponentially many Dirac delta peaks. In order to smooth the spectrum we shall average over a small frequency window $\d\om$, 
\ie we compute the EPR spectrum as 
\begin{equation}
 \tilde f(\om)=\frac1N\sum_{i=1}^N \frac{1}{\d\om}\int_\om^{\om+\d\om} f_i\ ,
 \quad f_i(\om)=\Re\argc{\int_0^\io\frac{\dd \t}{\p}\,g_i(\t)e^{-i\om\t-\eta\t}}
\end{equation}
where we introduce a small cutoff $\eta>0$. In essence we have to regularize integrals using a Cauchy principal value over finite intervals of length $\d\om$. We indeed note that 
\begin{equation}
 \int_0^\io\frac{\dd \t}{\p}\,e^{-i(\omega-\omega_0)\t-\eta\t}=\frac1\p\frac{1}{i(\omega-\omega_0)+\eta}=\frac{\eta/\p}{(\omega-\omega_0)^2+\eta^2}-\frac{i}{\p}\frac{\omega-\omega_0}{(\omega-\omega_0)^2+\eta^2}
\end{equation}
The real part is a Lorentzian and will act as $\d(\omega-\omega_0)$ for $\eta\to0^+$ while the other (principal value) imaginary part remains integrable around $\omega=\omega_0$ $\forall\eta>0$ (with a slope $\propto\eta^{-2}$).

 Performing the Fourier transforms yields
\begin{equation}\label{eq:finter}
 \begin{split}
  \tilde f(\om)=\frac{1}{N\d\om}\sum_{k,l,m,n}\Re&\left\{\argc{\mathbb{1}\argp{\om_{klmn}\in I_\om}-\frac{i}{\p}\ln\abs{\frac{\om+\d\om-\om_{klmn}}{\om-\om_{klmn}}}}\right.\\
  &\times\left. \r_{nm}\sum_i\bra{m}1-2iS_i^x\ketbra{k}{k}S_i^+\ketbra{l}{l}1+2iS_i^x\ket{n}\right\}
 \end{split}
\end{equation}
where $I_\om=[\om,\om+\d\om]$ and $\om_{klmn}=\varepsilon_k-\varepsilon_l+\mw(s^z_m-s^z_n)$. The index function $\mathbbm{1}$ is 1 if true and 0 else. The interval of EPR frequencies experimentally probed is centered around $\ome$. 
So, as the energy scales between different polarization sectors are well separated, the terms between brackets in the first line are 
negligible unless 
$s^z_k-s^z_l+s^z_m-s^z_n=1$. Yet the factor $\bra{k}S_i^+\ket{l}$ implies $s^z_k-s^z_l=1$, thus considering only matrix elements such that $s^z_m=s^z_n$ is enough to compute the EPR spectrum.

This means in the second line of Eq.~\eqref{eq:finter} the polarization from $\ket n$ to $\ket m$ must not change. As $S_i^+$ performs polarization jumps of $+1$, we can get rid of the non-polarization-conserving terms and get, relabeling indices, 
\begin{equation}\label{eq:tildefexpr}
\begin{split}
  \tilde f(\om)
  =\frac{1}{N\d\om}\sum_{n,k}\Re &  \left\{\argc{\mathbbm{1}(\om_{nk}\in I_\om)-\frac{i}{\p}\ln\abs{\frac{\om+\d\om-\om_{nk}}{\om-\om_{nk}}}}\right.\\
  &\times\left.  \argc{\sum_{\substack{m\\s^z_m=s^z_n}}\r_{mn}\sum_i\bra{n}S_i^+\ketbra{k}{k}S_i^-\ket{m}
  -\sum_{\substack{m\\s^z_m=s^z_k}}\r_{km}\sum_i\bra{m}S_i^-\ketbra{n}{n}S_i^+\ket{k}}
  \right\}
\end{split}
\end{equation}
Note that if the density matrix is diagonal in the eigenbasis, this implies
\begin{equation}
  \tilde f(\om)=\frac{1}{N\d\om}\sum_{n,k}\mathbbm{1}(\om_{nk}\in I_\om)(\r_{nn}-\r_{kk})\sum_i\abs{\bra{n}S_i^+\ket{k}}^2
\end{equation}
which agrees with the result of~\cite[Eq.(23)]{CFCRDL16}. In this case we have, noting $f(\om)=\sum_i f_i(\om)/N$ (for $\eta\to0^+$):
\begin{equation}\label{eq:Caraccioloetal}
 f(\om)=\frac{1}{N}\sum_i\sum_{n,m}\d(\om-\om_{nm})(\r_{nn}-\r_{mm})\abs{\bra{n}S_i^+\ket{m}}^2
\end{equation}
and only the gaps $\om_{nm}\approx\ome$ with $s_n^z=s_m^z+1$ actually contribute to the sum. So writing this set as an approximate interval (for large $N$) $I=\{\om_{nm}$ with $s_n^z=s_m^z+1\}$ where the EPR spectrum is non zero, one gets, using $\argc{S_i^+,S_i^-}=2S_i^z$,
\begin{equation}\label{eq:intEPR}
 \int_I \dd\om\, f(\om) = \frac1N\sum_i\argc{\sum_n\r_{nn}\bra{n}S_i^+S_i^-\ket{n}-\sum_m\r_{mm}\bra{m}S_i^-S_i^+\ket{m}}=\frac{2}{N}\sum_i\Tr(S_i^z\r)=\frac{2\moy{S^z}}{N}=P^z
\end{equation}
\ie the integral of the EPR spectrum gives the total polarization along $z$, $P^z$.

Finally, for the more trivial case of non-interacting spins\footnote{Note however that there the gaps are largely 
degenerated and one must take this into account in our dynamical equations.} with Hamiltonian 
$H_{\rm S}=\sum_i(\ome+\D_i)S_i^z=\sum_i\om_i S_i^z$, the previous integral holds as it is a particular case, 
but more precisely in Eq.~\eqref{eq:Caraccioloetal} at fixed spin $i$ the transition $\bra{n}S_i^+\ket{m}$ is non-zero 
iff $\om_{nm}=\om_i$, so that we can factor out the Dirac delta function and get
\begin{equation}
 f(\om)=\frac1N\sum_i\d(\om-\om_i)\sum_{n,m}(\r_{nn}- \r_{mm})\abs{\bra{n}S_i^+\ket{m}}^2
 =\frac1N\sum_iP_i^z\d(\om-\om_i)
\end{equation}
where $P_i^z=2\Tr(S_i^z\r)$. This last equation exhibits a connection between EPR and polarization profiles for non-interacting spins. 

\section{Bath-induced localization by temperature variation: additional numerical results}\label{sec:numerics}

\subsection{Thermal mixing and spin temperature behavior}~\label{sub:spinT}

It has been shown in Refs.~\cite{DLR15,DLRAMR16,RAMRDL18} within the secular GKLS equation and the Hilbert approximation that there exists a remarkable situation of \textit{thermal mixing} where the driven system behaves as an effective equilibrium steady state. It can be thought as an effective Gibbs ensemble with two parameters conjugated to the two conserved quantities of the isolated (microcanonical) system with Hamiltonian $H_{\rm S}$, which are the energy and the total spin along $z$. We note $\r^{\rm stat}$ the steady-state distribution, which is then diagonal in the eigenstate basis:
\begin{equation}
 \r^{\rm stat}_{nn}\propto e^{-\b_S(\varepsilon_n-hs^z_n)}
\end{equation}
where $\b_S$ is the so-called \textit{spin temperature} conjugated to the energy and $h$ is the effective magnetic field conjugated to the spin along $z$.


In the Hilbert approximation Eq.~\eqref{eq:hilbsec}, the transition rates define several timescales. 
In the following for simplicity we assume that the two $S_i^z$ transitions timescales are equal, and we note their common value $T_z=T(0)=T^*$. We shall comment later what happens when they differ. 
$T_1=2T(\ome)$ is the relaxation time of the system\footnote{The factor 2 is conventionally defined to recover the Bloch solution~\cite{B46} by solving exactly the single-spin case $N=1$ with $H_{\rm S}=\ome S_1^z$, 
\begin{equation*}
 \Tr\argp{ S_1^z \r^{\rm stat}}=\frac{M_0}{1+\frac{T_1}{T_{\rm MW}}}\ ,
\end{equation*}
~~\\
\begin{equation*}
\begin{split}
T_{\rm MW}=&\frac{1+(\ome-\mw)^2 T_2^2}{T_2\om_1^2}\ ,\\
   \frac{1}{T_2}=&\frac{1}{4T(0)}+\frac{1}{4T(\om_{\rm e})}\ ,\\
   M_0=&-\frac12\tanh\argp{\frac{\b\om_{\rm e}}{2}}=\frac{\Tr\argp{ S_1^z e^{-\b \ome S_1^z}}}{\Tr \,e^{-\b \ome S_1^z}}\ .
\end{split}
\end{equation*}}, and therefore can be considered as the longest timescale.
If we assume $T_z\ll T_1$ then the main timescales are:
\begin{itemize}
 \item $T_1$ the bath relaxation timescale induced by $S_i^x$ or $S_i^y$ flipping of the spins with associated energy $\om\approx\pm\ome$. At $T=1.2$ K these flips strongly favor the approximate ground state of all spins down: $h(\om\simeq\ome)\simeq 0.98$ and $h(\om\simeq-\ome)\approx 0.02$.
 \item $T_z$ the bath relaxation timescale due to $S_i^z$ flips with energy $|\om_{nm}|\ll\ome$. Such flips do not particularly favor any sign of the spin $i$, as $h(\om_{nm})\simeq1/2$ is quasi independent of the transition $\om_{nm}$.
 \item The microwave time 
 \begin{equation}
  \frac{1}{W_{nm}^{\rm MW}}=\frac{1+\argp{T_{nm}\D\om_{nm}}^2}{2\om_1^2T_{nm}\abs{\bra{n}S^x\ket{m}}^2}
 \end{equation}
 defined in Eq.~\eqref{eq:WMW}. These timescales depend on $n$ and $m$ but the shortest ones (\ie most efficient transitions) are such that $\abs{\bra{n}S^x\ket{m}}$ is not too low and $\D\om_{nm}\simeq0$ is near resonance. Therefore, as $T_{nm}\sim T_z$, this time may be assessed through a simple Lorentzian rate 
 \begin{equation}
   \frac{1}{T^{\rm MW}(\D\om)}=\frac{\om_1^2T_z}{1+(\D\om T_z)^2}
 \end{equation}
 of width $\D\om\sim 1/T_z$ and minimal typical value $T^{\rm MW}_{\rm min}=(\om_1^2T_z)^{-1}$.
 \item The Thouless time $T_{\rm D}$ is the typical dephasing timescale $1/\D\om_{nm}$ in the coherence dynamics 
 Eq.~\eqref{eq:secproj}, which is roughly given by the interaction and disorder timescales $T_{\rm D}=\min(1/U, 1/\D\ome)$.
\end{itemize}

This phase is characterized by an inhomogeneity of the polarization profile between the spins when the spin temperature is low. In a nutshell this occurs when the microwaves are effective in a short frequency range, which thus resonate with a small number of transitions $\om_{nm}$. Consequently they affect a small number of spins. But as shown in Refs.~\cite{DLR15,DLRAMR16,RAMRDL18}, thermal mixing happens when  the interactions strong enough, typically where the system Hamiltonian satisfies ETH. The spins then share the polarization and arrange into a charactistic polarization profile, characterized by an approximately linear relation $P(\om_i)$ with $\om_i=\ome+\D_i$ the typical frequency of spin $i$.

A difficulty in observing the spin-temperature behavior is that the same timescale $T_z$ controls several competing physical processes. It is the main timescale in the nonsecular terms for $T_z\ll T_1$. In the secular terms~\eqref{eq:secproj}, it \textit{(i)} enters in the microwave timescales as is clearly visible in the Hilbert approximation~\eqref{eq:hilbsec} 
\textit{(ii)} it enters the evolution of the diagonal elements, inducing the homogenization of the polarization profile as these $S^z$ transitions flip equally spins $+$ or $-$ (for this reason it is called \textit{spectral diffusion}) 
\textit{(iii)} participates out of the diagonal in the decoherence process through $T_{nm}$. 

Within the nonsecular GKSL equation~\eqref{eq:Lindrot}, we are able to recover a 
marked spin-temperature behavior under the following necessary conditions:
\begin{enumerate}
\item ~~
\begin{equation}
       \D\ome \lesssim U
      \end{equation}
meaning that the spin-temperature phase needs enough interactions for ETH to hold. 
Note that especially if $N$ is small the inequality is not sharp.
\item ~~
\begin{equation}
         T_z \gg T_{\rm D}\ (\textrm{Thouless time})\ \simeq\min\argp{\frac{1}{U},\frac{1}{\D\ome}} 
       \end{equation}
 In the coherence dynamics, the dephasing dominate the nonsecular terms and the  Hilbert approximation is  valid.
\item ~~
\begin{equation}\label{eq:MWeffective}
         T_1T_z\om_1^2\gtrsim1 \qquad\Leftrightarrow\qquad T^{\rm MW}_{\rm min}\lesssim T_1
       \end{equation}
 The microwaves are effective at least for the resonant gaps (otherwise the system would thermalize with Boltzmann distribution).
\item ~~
\begin{equation}
        \om_1\sqrt{\frac{T_1}{T_z}}\lesssim\max(U, \D\ome)
\end{equation}
 $\om_1\sqrt{T_1/T_z}$ is the maximal frequency range for which $T^{\rm MW}(\D\om)<T_1$ under the previous condition Eq.~\eqref{eq:MWeffective}, \ie the frequency window over which the microwaves are effective. This range needs to be narrower than the typical frequency range $\max(U, \D\ome)$ which is probed in the EPR experiment (\ie frequencies around $\ome$ which correspond to gaps of $S_i^+$). Otherwise, most transitions are irradiated and all spins feel the microwaves, which has a tendency to thermalize them at 
 infinite temperature (\ie $\Tr (S_i^z\r_{\rm stat}) =0$). This is needed to create some inhomogeneity within the spins. 
 \item ~~
 \begin{equation}\label{eq:TzT1}
        T_z\lesssim T_1
       \end{equation}
This emphasizes that if $T_1$ is too large then spectral diffusion will dominate and the profile will be homogeneous (high spin temperature). 
A way to keep spectral diffusion low is to make the two timescale differ such that
\begin{equation}
        T(0)\ll T^*
\end{equation}
Indeed $T(0)$ enters only in $T_{nm}$~\eqref{eq:bathrate}, \ie in the decoherence process off the diagonal of Eq.~\eqref{eq:secproj}, while $T^*$ appears in the bath transitions~\eqref{eq:bathrate} on the diagonal which determine directly the steady-state values.          
  \item \begin{equation}
        \om_1T_z\gtrsim 1 \qquad\Leftrightarrow\qquad T^{\rm MW}_{\rm min}\lesssim T_z
        \end{equation}
        This is again to prevent spectral diffusion to destroy the spin-temperature behavior. 
\end{enumerate}


\begin{figure*}[t]
 \begin{tabular}{cc}
  \includegraphics[width=0.45\textwidth]{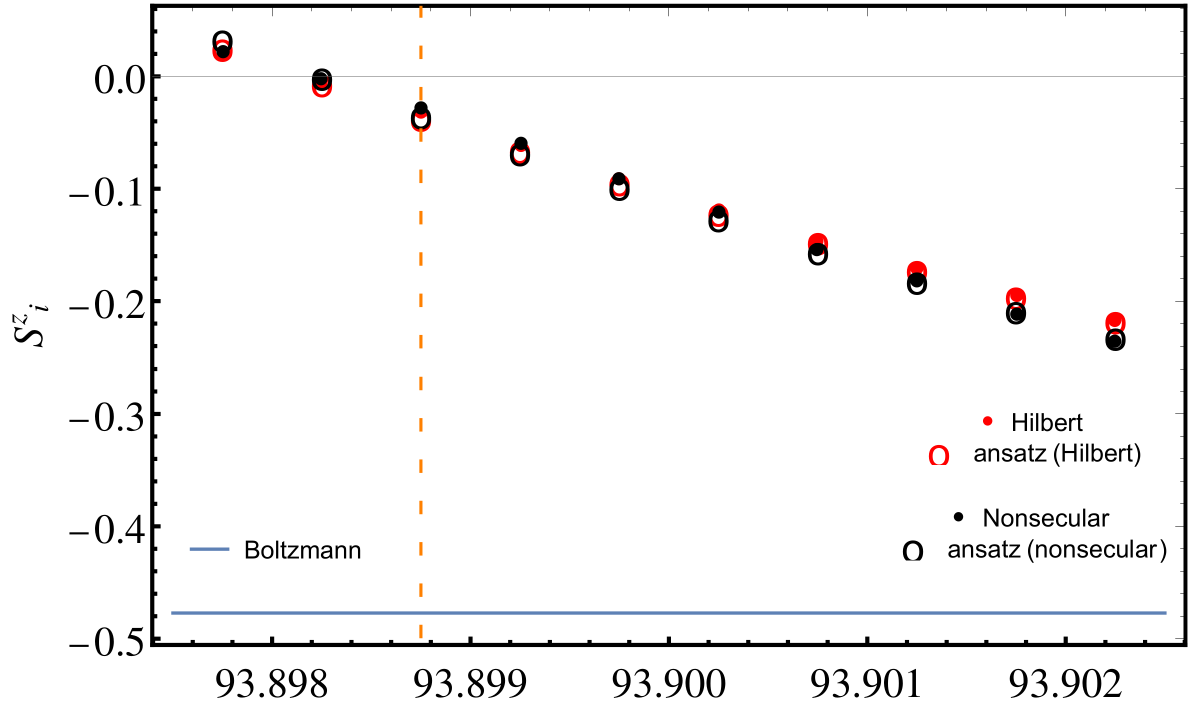} &
  \includegraphics[width=0.45\textwidth]{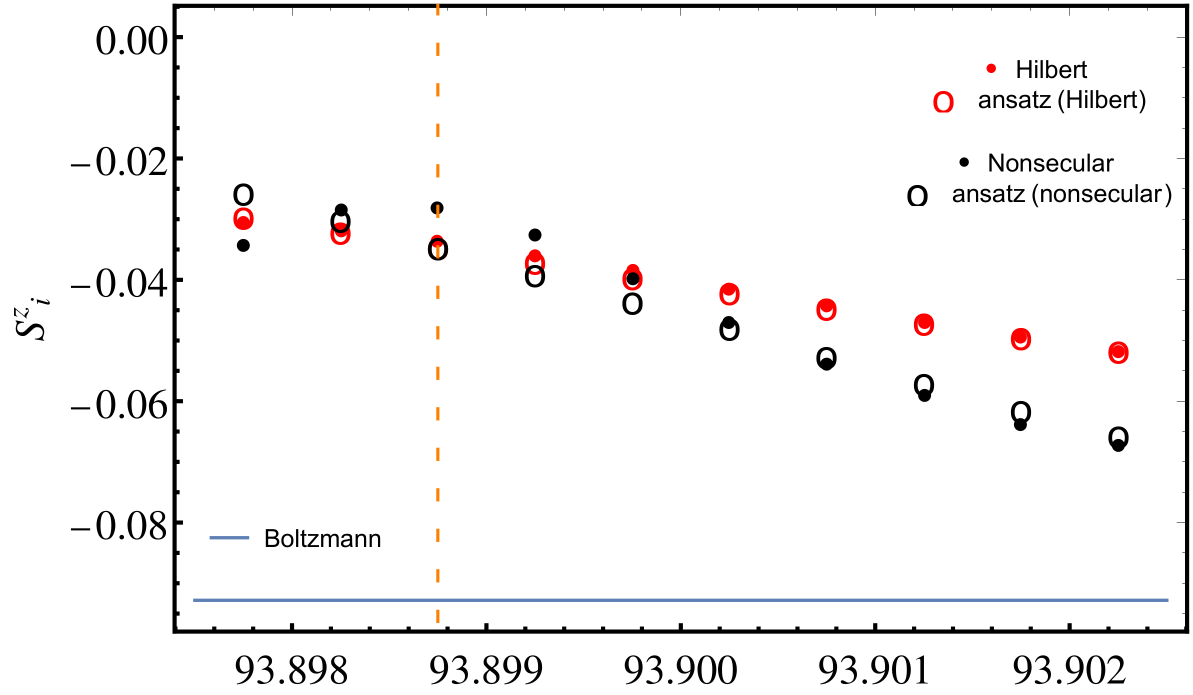} \\
  \hspace{1cm}$\ome+\D_i$ $(2\p$GHz) & \hspace{1cm}$\ome+\D_i$ $(2\p$GHz)
 \end{tabular}
\caption{\label{fig:ETHpol} Polarization profiles (red : Hilbert, black : nonsecular) for $N=10$ spins with  $\D\ome=5\cdot2\p$MHz and $U=0.75\cdot2\p$MHz (ETH eigenstates). 
The bath and microwave parameters are given in Table I of the main text ($T_1=10^4(2\p\mathrm{GHz})^{-1}$, $T^*=T(0)=10^3(2\p\mathrm{GHz})^{-1}$). 
The vertical dashed line points out the microwave frequency. 
The horizontal blue line is the Boltzmann prediction. 
Full dots are the numerical results~; hollow circles are obtained by the spin-temperature ansatz~\eqref{eq:spinTans}. 
(left) $\b^{-1}=1.2$ K. The bath is slow and as a consequence nonsecular and Hilbert dynamics yield similar spin-temperature curves.
(right) $\b^{-1}=12$ K.  Here the bath timescales are short and 
the full (nonsecular) dynamics gets localized : only the near-resonant spins feel the microwaves, as if they were non interacting, causing the hole burning. 
Each profile has been averaged over 1000 realizations of disorder and interactions for the nonsecular case and 3000 realizations for the Hilbert case.}
\end{figure*}

The spin-temperature shapes of the EPR spectrum are displayed in main text's Fig. 2(top). 
In Fig.~\ref{fig:ETHpol} we show the corresponding polarization profiles in the ETH phase. 

\subsection{Estimating the spin-temperature ansatz parameters from simulation data}\label{sub:fit}

Here we give details about the numerical computation of the spin-temperature ansatz parameters. 
The spin-temperature ansatz is
\begin{equation}\label{eq:spinTans}
 \r_{nn}^{\rm ans}(\b_s,h)\propto e^{-\b_s(\varepsilon_n-hs_n^z)}
\end{equation}
We solve numerically for $(\b_s,h)$ to match the average energy and polarization in the stationary state~\cite{DLRAMR16}
\begin{equation}
 \left\{ 
 \begin{split}
  \Tr(H_{\rm S}\r^{\rm stat})=&\sum_{n=1}^{2^N}\varepsilon_n\r_{nn}^{\rm ans}(\b_s,h)\\
  \Tr(S^z\r^{\rm stat})=&\sum_{n=1}^{2^N}s^z_n\r_{nn}^{\rm ans}(\b_s,h)
 \end{split}\right.
\end{equation}
through an iterative procedure, starting from the initial guess
\begin{equation}
  \left\{ 
 \begin{split}
 \b_s=&\frac{2N}{\D\ome}\,\mathrm{argth}\,P_{\rm n}^{\rm stat}\\
 h=&\mw
 \end{split}\right.
\end{equation}
where the nucleus polarization is~\cite{CSFCRT14}
\begin{equation}
 P_{\rm n}^{\rm stat}=\frac{\int \dd\om\, d(\om)d(\om+\om_{\rm n})[P^z_{\rm stat}(\om)-P^z_{\rm stat}(\om+\om_{\rm n})]}{\int \dd\om\, d(\om)d(\om+\om_{\rm n})[1-P^z_{\rm stat}(\om)P^z_{\rm stat}(\om+\om_{\rm n})]}
 \simeq\frac{2\sum_{i=1}^{N-1}\argc{\Tr(S_i^z\r^{\rm stat})-\Tr(S_{i+1}^z\r^{\rm stat})}}{\sum_{i=1}^{N-1}\argc{1-4\Tr(S_i^z\r^{\rm stat})\Tr(S_{i+1}^z\r^{\rm stat})}}
\end{equation}
$\om_{\rm n}$ is the nucleus Zeeman gap. $d(\om)$ is the disorder distribution at energy $\om$ ; the second equality comes from considering the disorder as uniform and the typical frequency between two spins $\om_i-\om_{i+1}\approx\om_{\rm n}$, where $\om_i=\ome+\D_i$ is the frequency of spin $i$ in a non-interacting picture. We finally check that for all spins $\Tr(S_i^z\r^{\rm ans})=\Tr(S_i^z\r^{\rm stat})$ in order to control that the ansatz reproduces well the data.

 Here the polarization of a coupled nucleus enters explicitly. Note that in this work we focused on the electronic spins only, as their out-of-equilibrium steady state is the crucial feature concerning DNP in the thermal mixing regime. The electronic polarization is indeed transferred to the nuclei, the polarization of interest in DNP \textit{in fine}. One can actually add a nucleus to the spin system to check the polarization transfer; this has been studied in Ref.~\cite{DLRAMR16}.

\subsection{Hole burning}\label{sec:HB}

A very different situation where spins do not have collective behavior is for example when the disorder is strong~\cite{DLR15,DLRAMR16,RAMRDL18}, in the phase where the eigenstates are many-body localized. Then only spins resonant with the microwave feel them and acquire a vanishing polarization (\ie a large local  temperature, spins are heated up), while the others are unaffected by the driving (and are close to be all down spins as bath temperature is low), as if they were independent spins. The polarization profile results in a hole burning shape. 
We showed in this article that if the coupling to the bath is strong enough, the nonsecular terms may induce this hole burning shape as well in the ETH phase, as if spins were localized by the bath interaction. 

The necessary conditions for this bath-induced localization to take place are 1, 3 and 4 of Sec.~\ref{sub:spinT} but now the nonsecular 
terms need to be as important as possible to dominate over dephasing.
This means we take $T^*=T(0)=T_z$ with
\begin{equation}
     T_z \lesssim T_{\rm D}\ (\textrm{Thouless time})\ \simeq\min\argp{\frac{1}{U},\frac{1}{\D\ome}}
\end{equation}
Note that the rate of the $1/T_{nm}$ decoherence term is as well controlled by $T_z$.

%

The hole burning at high temperature is visible in the ETH phase in Fig.~\ref{fig:ETHpol}(right) for the polarization and in Fig. 2(bottom) of the main text for the EPR spectrum. 
In Fig.~\ref{fig:MBL} we show that if the Hamiltonian eigenstates are MBL, one gets, independently of the chosen dynamics and temperatures, a hole burning behavior.

\begin{figure*}[h!]
 \begin{tabular}{cc}
 (a) & (c) \\
  \includegraphics[width=0.4\textwidth]{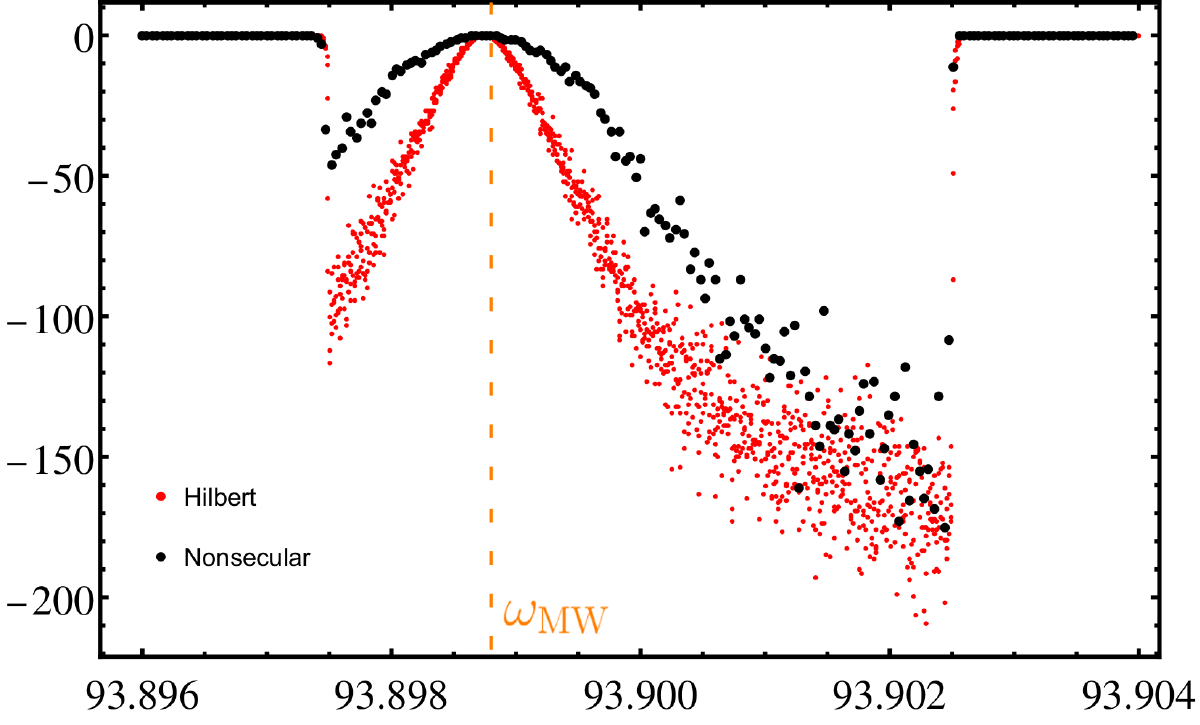} &
  \includegraphics[width=0.4\textwidth]{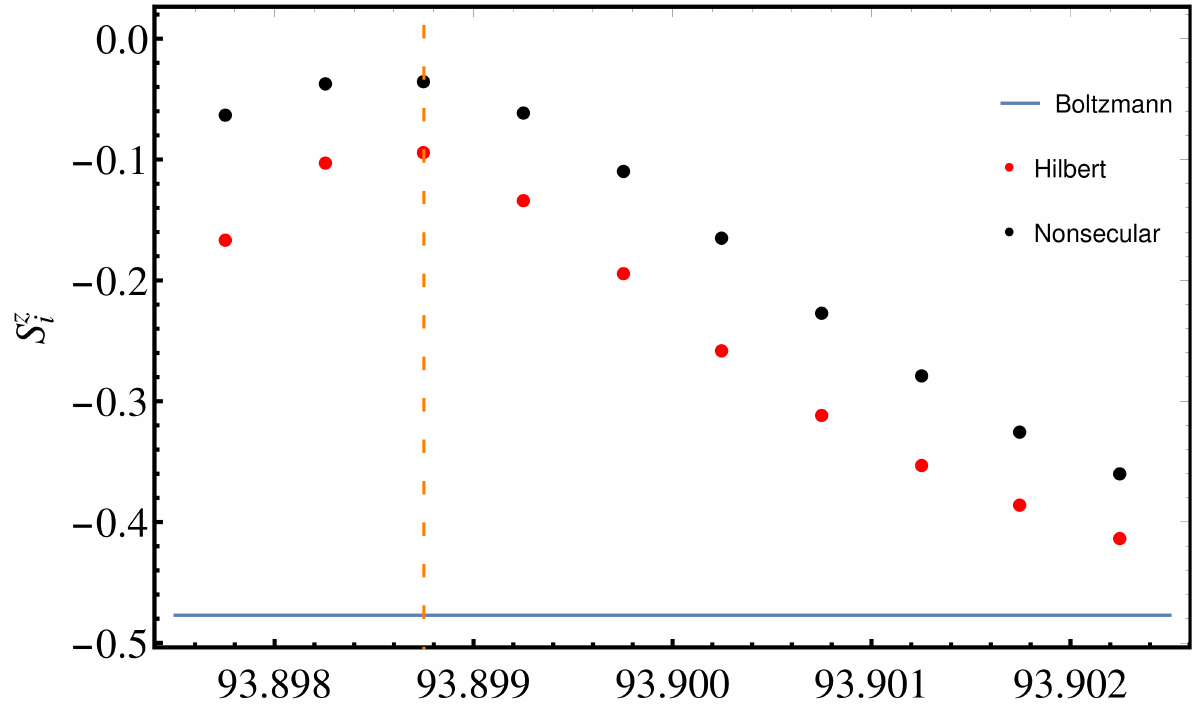} \\
    \includegraphics[width=0.4\textwidth]{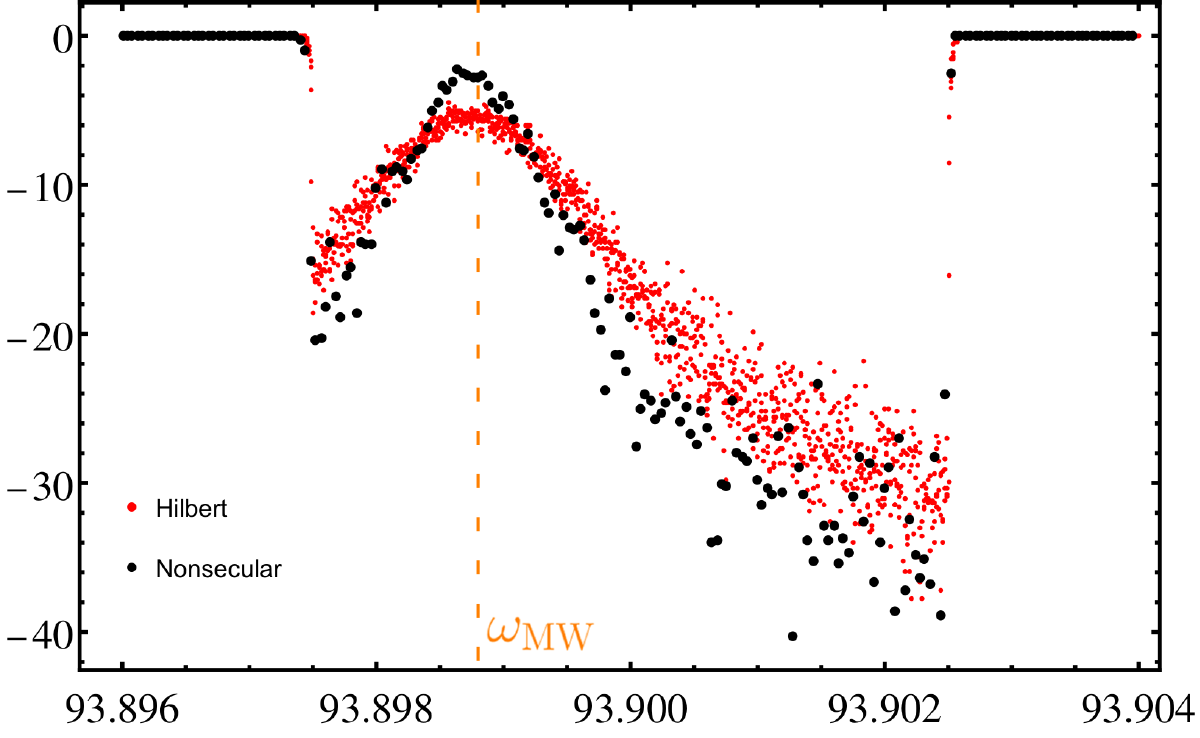} &
  \includegraphics[width=0.4\textwidth]{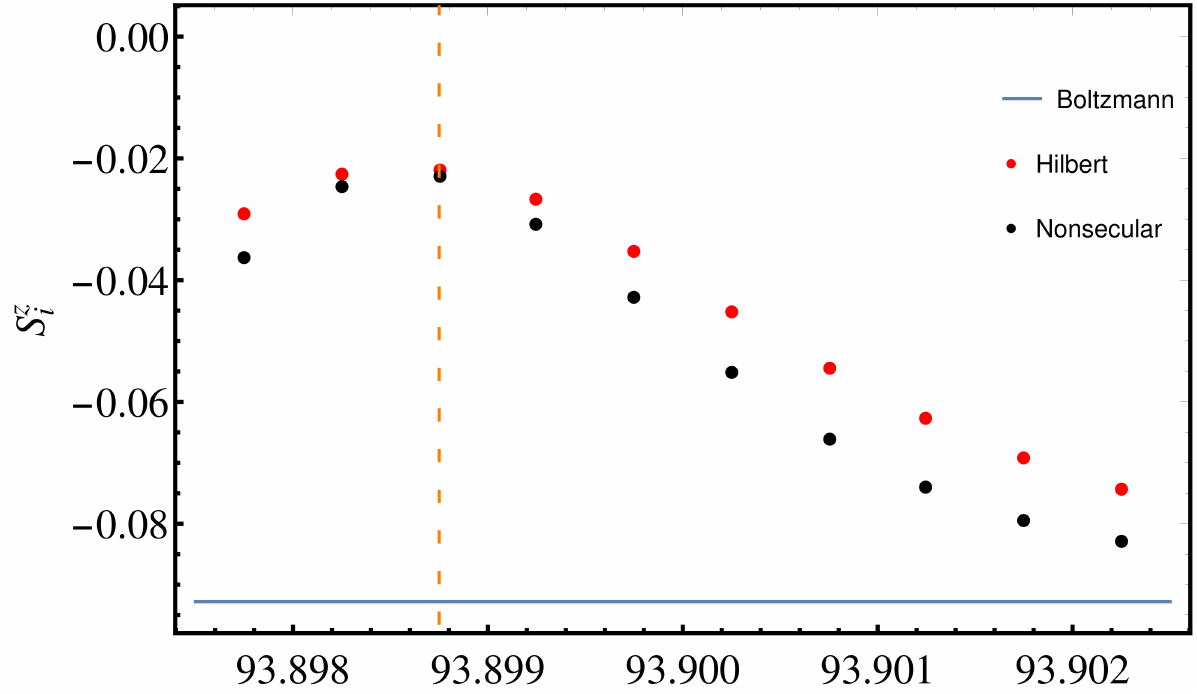} \\
  $\om$ $(2\p$GHz) & \hspace{1cm}$\ome+\D_i$ $(2\p$GHz)\\
  (b) & (d)
 \end{tabular}
\caption{\label{fig:MBL} 
Numerical profiles (red : Hilbert, black : nonsecular) for  $N=10$ spins with $\D\ome=5\cdot2\p$MHz and $U=0.1\cdot2\p$MHz (MBL eigenstates). 
The bath and microwave parameters for each temperature are given in main text's Table I.
The vertical dashed line points out the microwave frequency. 
Average is taken over 1000 realizations of disorder and interactions for the nonsecular case and
10000 realizations in the Hilbert case. 
$\b^{-1}=1.2$ K in the top figures ; $\b^{-1}=12$ K in the bottom figures.
(a)(b) EPR spectra. 
Frequency bins are ten times larger in the nonsecular case than in the Hilbert case.
(c)(d) Polarization profiles. The horizontal blue line is the Boltzmann prediction.\\ 
Here we note a hole burning shape for both temperatures. There is no qualitative effect of nonsecularity : nonsecular terms are small for MBL eigenstates 
as mentioned in Sec.~\ref{sub:comparison}.}
\end{figure*}

In the data exhibited in the main text we have chosen $T_1$ to be the longest timescale (the other timescale of the nonsecular terms being $T^*$). 
In Fig.~\ref{fig:T1small} we display the EPR and polarization profiles obtained for the same system as in Fig. 2(bottom) of the main text (see Fig.~\ref{fig:ETHpol}) but with $T_1\ll T^*$. We obtain again a breakdown of thermal mixing. 

\begin{figure}[h!]
 \begin{tabular}{cc}
  \includegraphics[width=0.45\textwidth]{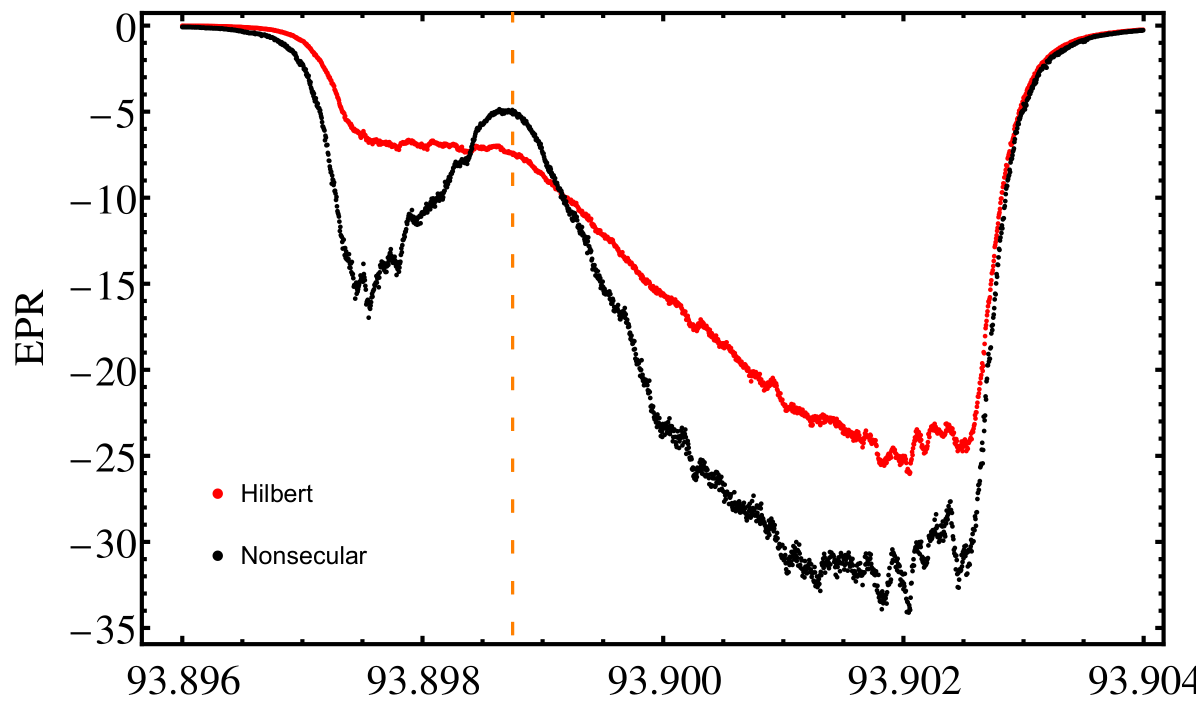} &
  \includegraphics[width=0.45\textwidth]{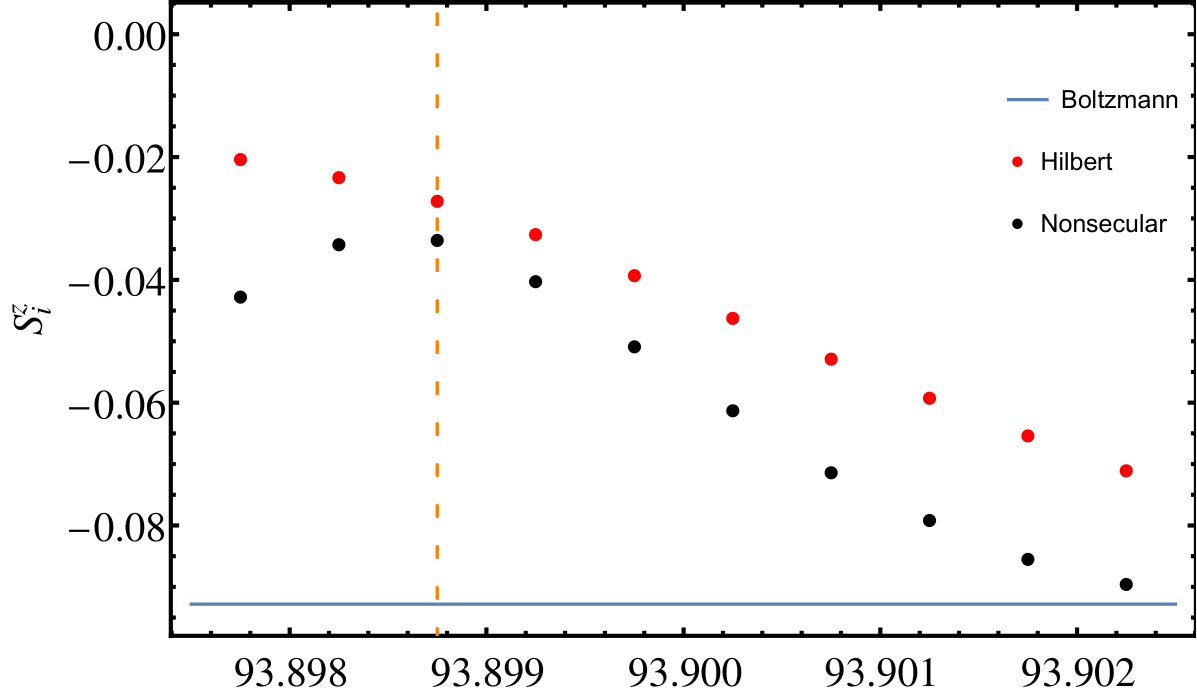} \\
  \hspace{1cm}$\om$ $(2\p$GHz) & \hspace{2cm}$\ome+\D_i$ $(2\p$GHz)
 \end{tabular}
\caption{\label{fig:T1small} Polarization profiles (red : Hilbert, black : nonsecular) for $N=10$ spins with  $\D\ome=5\cdot2\p$MHz and $U=0.75\cdot2\p$MHz (ETH eigenstates) $\b^{-1}=12$ K. 
The bath and microwave parameters are $T_1=10^4(2\p\mathrm{GHz})^{-1}=1.6\,\m\mathrm{s}$, $T^*=5\cdot10^5(2\p\mathrm{GHz})^{-1}=80\,\m\mathrm{s}$, $T(0)=2\cdot10^4(2\p\mathrm{GHz})^{-1}=3.2\,\m\mathrm{s}$. 
(left) EPR spectrum. (right) Polarization profiles. 
The vertical dashed line points out the microwave frequency. 
The horizontal blue line is the Boltzmann prediction. 
Each profile has been averaged over 1000 realizations of disorder and interactions for the nonsecular case and 3000 realizations for the Hilbert case.}
\end{figure}

In Fig.~\ref{fig:irrealistic} we show that one can get an even more impressive localization effect 
from the nonsecular GKSL equation without disorder (see the caption for the parameters). This is achieved for $N=12$ spins by pushing the bath relaxation times 
to less realistic values and irradiating only a single spin (namely, the third one). 
In the Hilbert approximation, the 12 spins share the same polarization owing to the strong interaction. All spins feel the microwaves.  
But in the nonsecular dynamics, despite  the strong interaction, the third spin is put to infinite temperature ($\Tr(S_3^z\r_{\rm stat})=0$) 
whereas all other spins are at Boltzmann equilibrium. 

\begin{figure}[h!]
\centering
\includegraphics[width=0.45\textwidth]{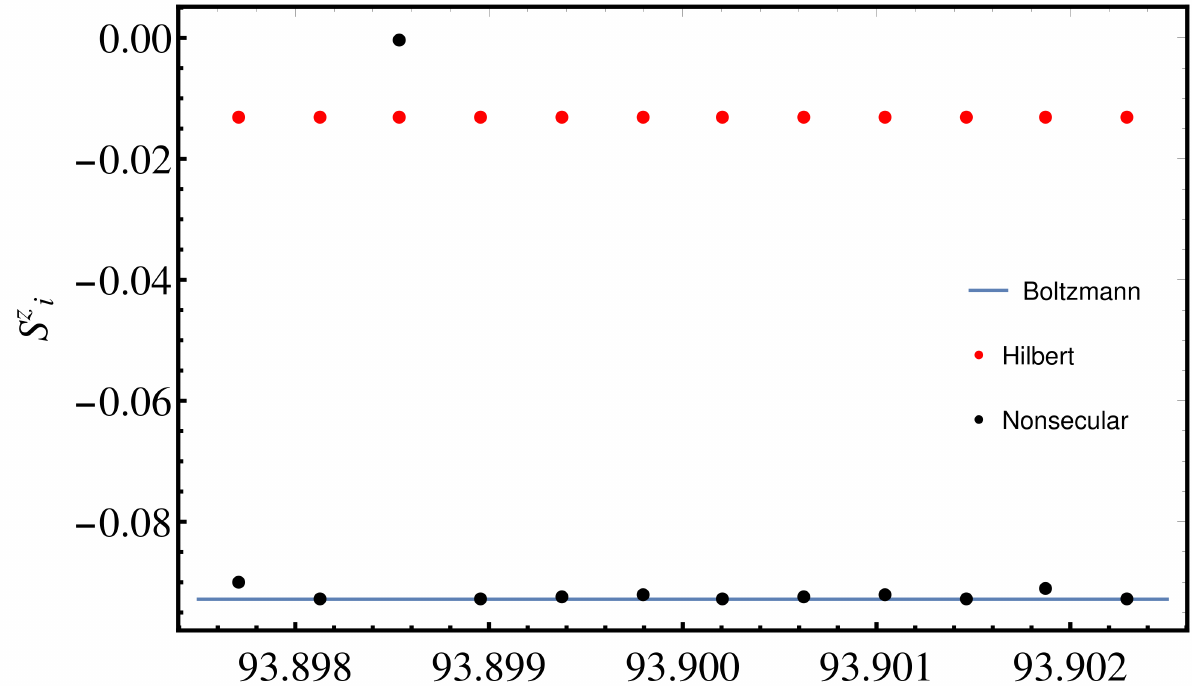} \\
  \hspace{2cm}$\ome+\D_i$ $(2\p$GHz)
\caption{\label{fig:irrealistic} Polarization profile for a setup exhibiting a very strong localization, for $N=12$ spins at $\b^{-1}=12$ K 
(red : Hilbert, black : nonsecular). The data corresponds to a single typical sample where disorder is negligible ($\D\ome=10^{-4}\cdot2\p$MHz, essentially to ensure 
that the gaps are non degenerate) and the interaction is $U=2\p$ MHz. 
The microwaves irradiate \textit{only} the third spin, with frequency $\mw=\ome$ and strong amplitude $\om_1=0.1\cdot2\p$GHz. 
The nonsecular terms are very effective due to the very low value of $T(0)=T^*=0.1\cdot(2\p\textrm{GHz})^{-1}=10^{-10}/2\p$ s ; $T_1=10^{-4}/2\p$ s. 
}
\end{figure}

\subsection{Spectral properties of the stationary state}\label{sub:spectral}

In this section we provide additional data concerning the spectral properties of the steady-state density matrix $\r_{\rm stat}$ 
in the Hilbert or nonsecular cases. 
We emphasize that, for all the $N=10$ data, the sample-to-sample fluctuations are quite well concentrated on the averages displayed. 

In Fig.~\ref{fig:10} we provide the eigenvalue distribution of the stationary density matrix, for parameters where the nonsecular 
dynamics leads to a hole burning while the Hilbert one yields a spin-temperature shape. 
This is to show that the distribution is not dominated by a small number of eigenstates ; all of them are relevant. Moreover, the  
distribution in both dynamics are similar : the qualitative difference in steady-state observables does not come from this population 
distribution but finds an explanation in the eigenstates' statistics, as explained in the main text.

\begin{figure*}[h!]
\centering
\includegraphics[width=0.45\textwidth]{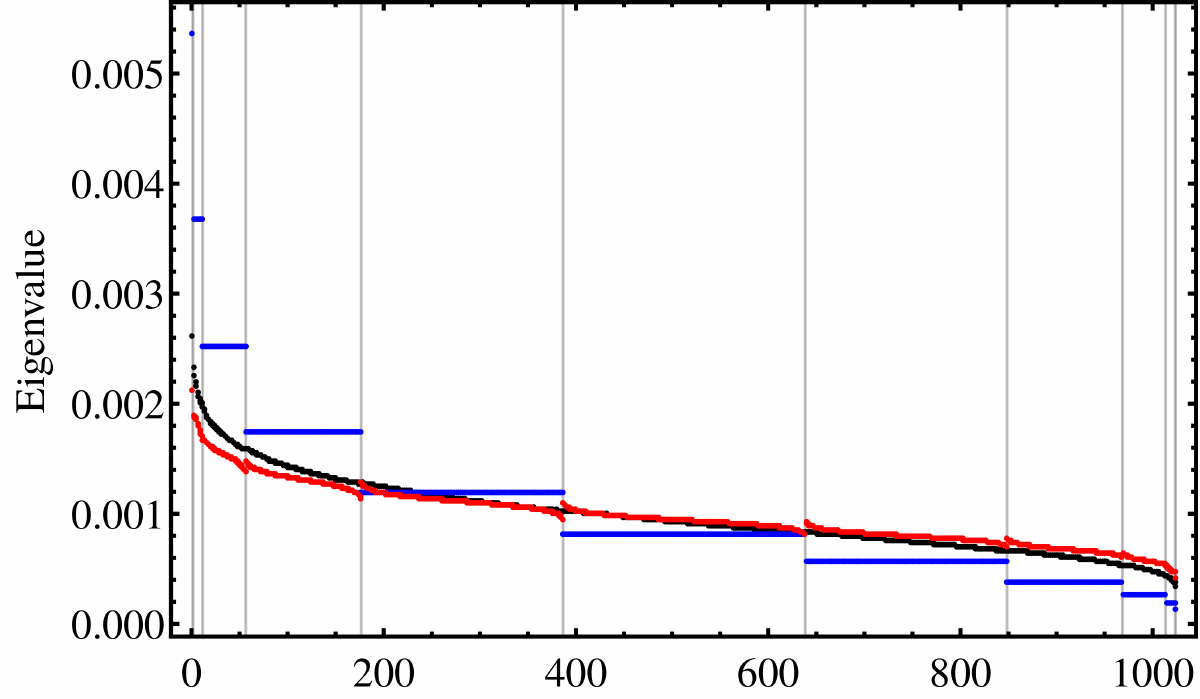} \\
  $\qquad\qquad \l$ 
\caption{\label{fig:10} 
Eigenvalues (populations) $\bra\l\r_{\rm stat}\ket \l$ of the stationary density matrix ($\ket\l$ are eigenvectors), 
classified on the horizontal axis by increasing energy $\bra\l H_{\rm S}\ket \l$. 
Parameters correspond to those of Fig.~\ref{fig:ETHpol} for  $N=10$ spins. 
Vertical gray lines delimit sectors of constant polarization (for Hamiltonian eigenstates). 
Blue : Boltzmann equilibrium $\r_{\rm stat}\propto \exp(-\b H_{\rm S})$. 
Red : Hilbert solution. For the latter cases, the eigenvectors $\ket\l=\ket n$ are those of the Hamiltonian. 
Black : nonsecular solution.}
\end{figure*}

It is pointed out in the main text that the when when the bath induces localization in the system, the eigenstates reached by the 
density matrix in the long-time limit are different from the Hamiltonian ones. They do not satisfy ETH,  are less entangled and more akin to MBL eigenstates. 
Now we focus on the more extreme case introduced at the end of Sec.~\ref{sec:HB}: 12 homogeneous spins, only spin 3 irradiated with strong 
nonsecular effects. Here the localization is ``complete'' : spin 3 decouples from the rest of the system, in spite of the strong 
interactions. 
In Fig.~\ref{fig:12} we demonstrate that the eigenstates of the density matrix is not anymore the ETH Hamiltonian eigenstates but 
are pure states factorized on each individual spin basis, \ie all $S_i^z$ are good quantum numbers to describe this eigenbasis ; it corresponds 
to the Hamiltonian basis for non-interacting spins $U=0$. Here the measurements performed by the bath modes wipe out 
all interaction in the many-body system. 
The entanglement entropy of each eigenstate is thus zero. The eigenvalue distribution of the matrix is much closer to the Boltzmann one 
than the Hilbert one.

\begin{figure*}[h!]
 \begin{tabular}{cc}
 (a) & (c) \\
  \includegraphics[width=0.45\textwidth]{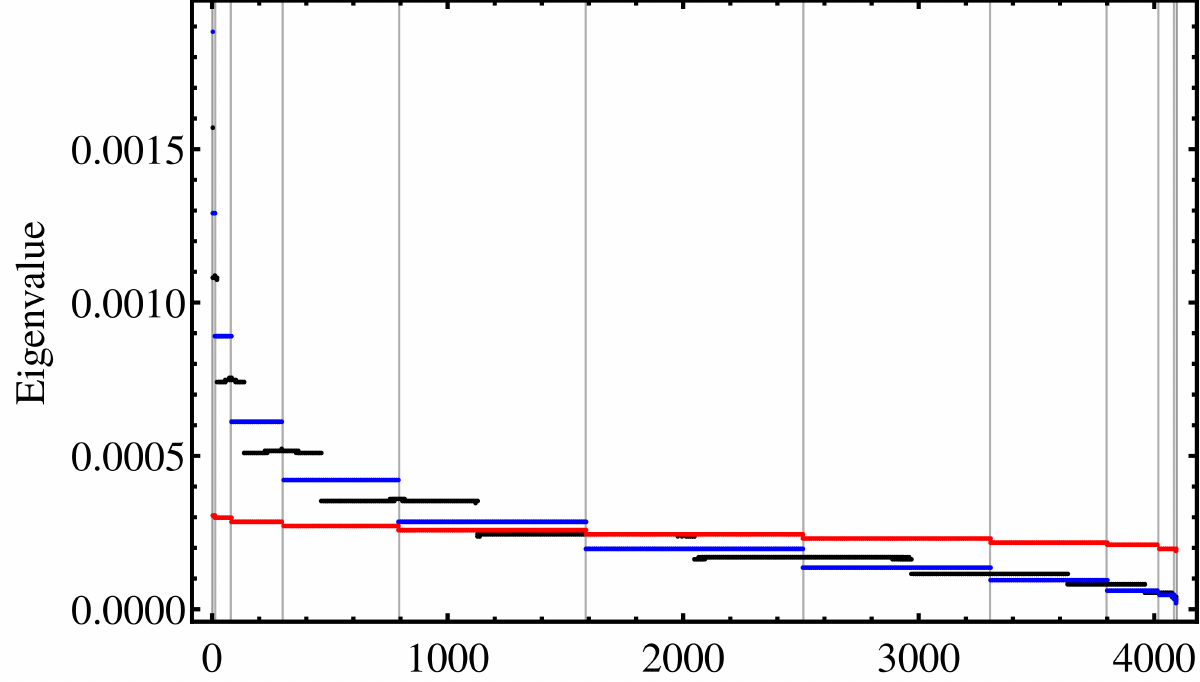} &
  \includegraphics[width=0.45\textwidth]{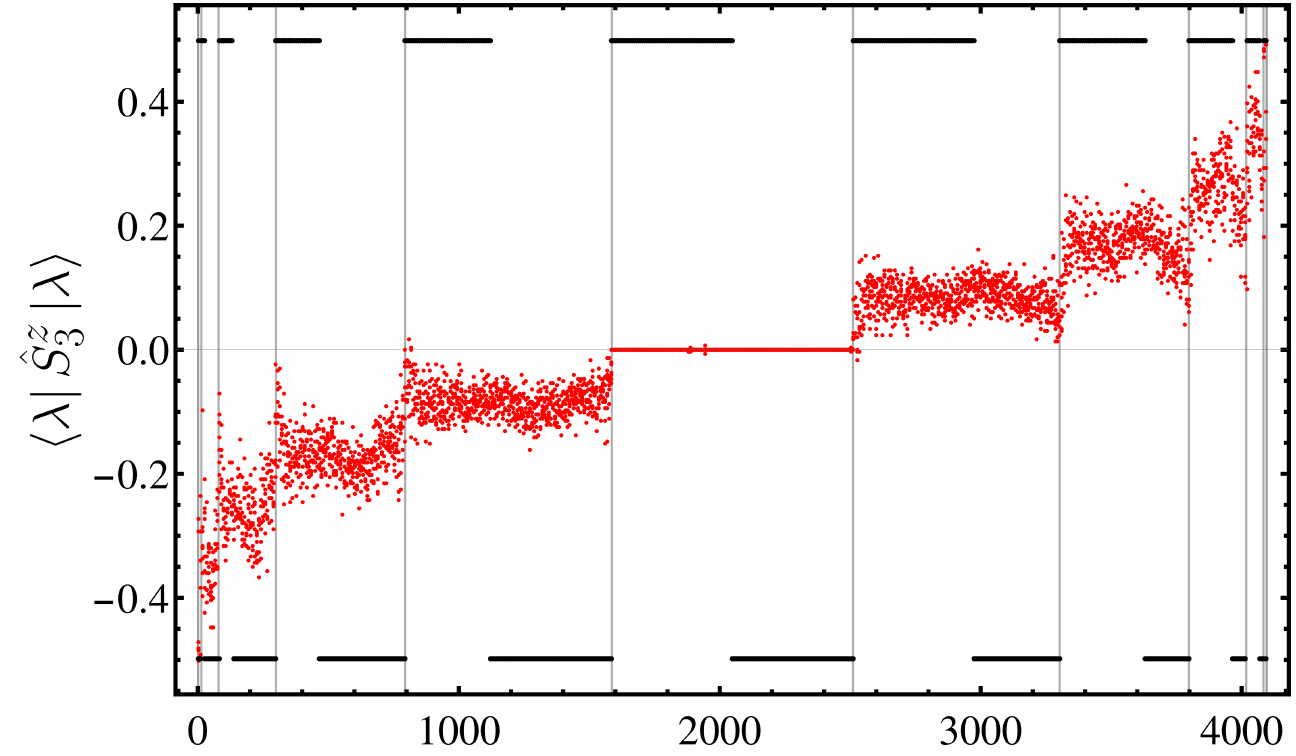} \\
    \includegraphics[width=0.45\textwidth]{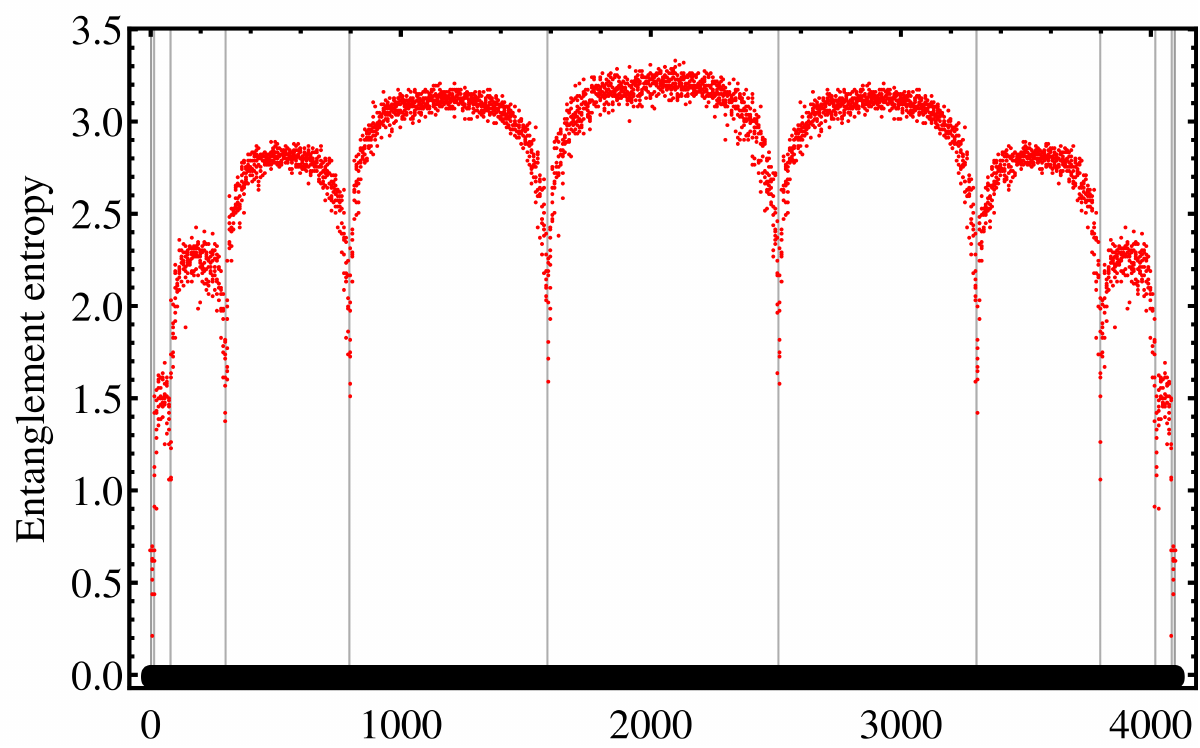} &
  \includegraphics[width=0.45\textwidth]{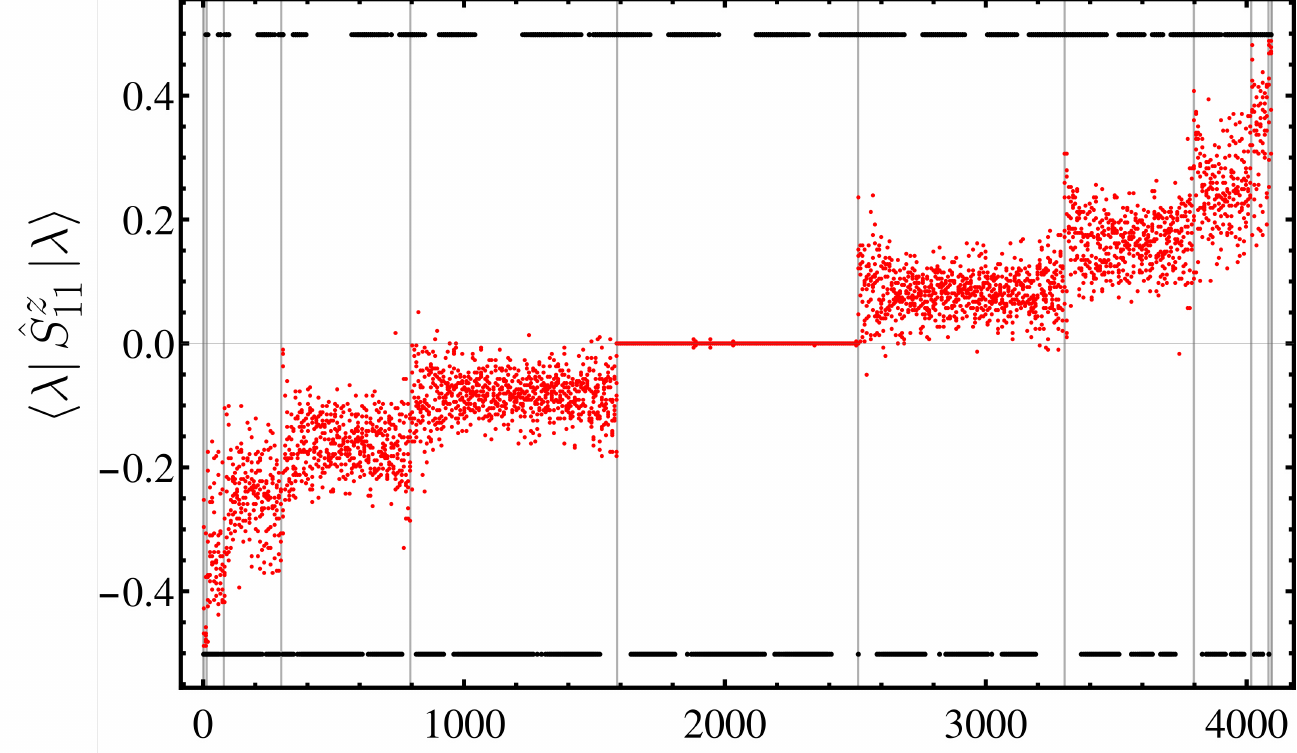} \\
$\qquad\qquad \l$ & $\qquad\qquad \l$\\
  (b) & (d)
 \end{tabular}
\caption{\label{fig:12} 
The setting corresponds to the fully localized one with 12 spins introduced at the end of Sec.~\ref{sec:HB}.
Index $\l$ is classified on the horizontal axis by increasing energy $\bra\l H_{\rm S}\ket \l$ ($\ket\l$ are eigenvectors of 
$\r_{\rm stat}$). 
Blue : Boltzmann equilibrium $\r_{\rm stat}\propto \exp(-\b H_{\rm S})$. 
Red : Hilbert solution. For the latter cases, the eigenvectors $\ket\l=\ket n$ are those of the Hamiltonian. 
Black : nonsecular solution. 
Vertical gray lines delimit sectors of constant polarization (for Hamiltonian eigenstates).
(a) Eigenvalues (populations) $\bra\l\r_{\rm stat}\ket \l$ of $\r_{\rm stat}$. 
(b) Entanglement entropy of each eigenvector. Partial trace is taken over spins 6 to 12. The black dots have zero entanglement entropy and are localized pure states of the $\arga{S_i^z}$ basis. 
(c) Expectation value of the observable $S_3^z$. 
(d) Same for $S_{11}^z$, qualitatively close to any other spin.
 }
\end{figure*}

%
%
\newpage

\section{A toy model for Zeno localization}\label{sec:3levels}

\begin{figure}[b]
 \includegraphics[width=0.5\linewidth]{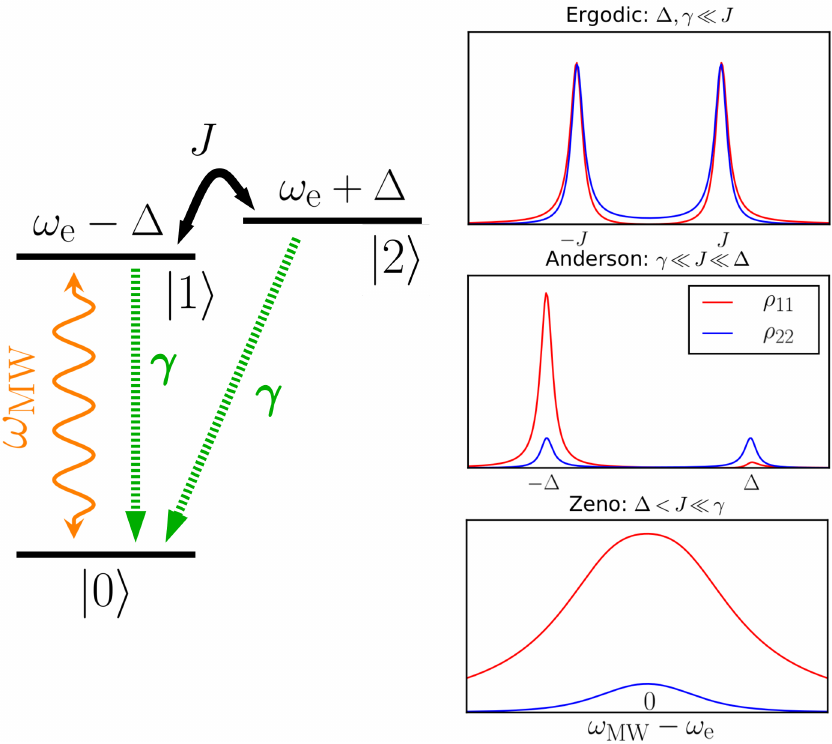}
 \caption{(left) A system of three levels $|0\rangle,|1\rangle,|2\rangle$ with energies 
 $0,\omega_\mathrm{e}\pm\Delta$ and a coupling $J$ between $|1\rangle$ and $|2\rangle$.
 A monochromatic drive at frequency~$\mw$ couples $|0\rangle$ and $|1\rangle$. 
 Two independent zero-temperature baths lead to decay of the population from  $|1\rangle$ and $|2\rangle$ to $|0\rangle$ with  rate $2\gamma$. The largest energy scale is~$\ome$. 
(right) Level populations $\rho_{ii}(\mw-\ome)$ in three limits. 
 In the Anderson limit the strong imbalance between $\rho_{11}$ and $\rho_{22}$ is caused by the eigenstates' localization on the respective levels. In the Zeno limit the system eigenstates are uniformly spread over $|1\rangle$ and $|2\rangle$, but they do not have time to
 form because of the dissipation.}
 \label{fig:3levels}
\end{figure}

The essence of the difference between the Zeno and the Anderson localization in a driven-dissipative system can be illustrated by a simple example with just three levels. In this section, we analyze such a toy model, sketched in Fig.~\ref{fig:3levels}. It presents similar features to the many-body DNP model with the advantage of possessing an analytical solution. 
We put hats on some operators to avoid possible confusion with scalar quantities.

Let us consider a three-level system (states labeled 0, 1, 2) subject to a monochromatic microwave perturbation that only couples the ground state $\ket 0$ and the state $\ket1$. The system is coupled to two uncorrelated harmonic baths that allow exchange of energy between the ground state and the excited states. The Hamiltonian reads
\begin{eqnarray}
&& {H}= {H}_\mathrm{S}+ {H}_\mathrm{MW}(t)
+ {H}_\mathrm{int}+ {H}_\mathrm{B}={}\nonumber\\
&&\quad{}=\left(\begin{array}{ccc}
0 & V^*e^{i\mw{t}}+\hat {B}_1^\dagger &\hat  {B}_2^\dagger \\
Ve^{-i\mw{t}}+\hat {B}_1 & \ome-\Delta & J \\ 
 \hat{B}_2  & J & \ome +\Delta
\end{array}\right)
+\sum_{\alpha=1,2}\sum_k\Omega_k \hat{a}_{\alpha{k}}^\dagger \hat{a}_{\alpha{k}}\\
&& \hat{B}_\alpha=\sum_kg_{\alpha{k}} \hat{a}_{\alpha{k}}
\end{eqnarray}
The Born-Markov master equation~\eqref{eq:markov} (no secular approximation is involved) for the system density matrix $ \rho$ 
in the Schr\"odinger representation has the form
\begin{equation}
\frac{\dd \rho}{\dd t}=-i\left[ H_{\rm S}+ H_{\rm MW}, \rho\right]
-\int\limits_0^\infty\mathrm{d}\tau\,
\Tr_\mathrm{B}\left\{\left[ {H}_\mathrm{int},\left[
e^{-i( H_{\rm S}+ H_{\rm B})\tau} {H}_\mathrm{int}
e^{i( H_{\rm S}+ H_{\rm B})\tau}, \rho\otimes \rho_\mathrm{B}\right]\right]\right\}.
\end{equation}
It does not depend on the choice of basis in the system subspace and thus does not favor the eigenbasis of~$ H_{\rm S}$. 
Assuming $J,\Delta\ll\ome$, we approximate
\begin{equation}
 e^{-i( H_{\rm S}+ H_{\rm B})\tau} {H}_\mathrm{int}
e^{i( H_{\rm S}+ H_{\rm B})\tau}
\approx\sum_{\alpha=1,2}|\alpha\rangle\langle0|
\sum_kg_{\alpha{k}} \hat{a}_{\alpha{k}}e^{i(\Omega_k-\ome)\tau}+\ \mathrm{H.c.}\equiv
\sum_{\alpha=1,2}|\alpha\rangle\langle0|\, \hat{B}_\alpha(-\tau)\,
e^{-i\ome\tau}+\ \mathrm{H.c.}
\end{equation}
with $\hat{B}_\alpha(\tau)=\sum_kg_{\alpha{k}} e^{-i\Omega_k\tau}\hat{a}_{\alpha{k}}$ the dynamical evolution of the bath modes. 
When performing the trace over the bath modes, assuming the bath has no fine structure on the scale~$J,\Delta$, we must only evaluate the following bath correlators at frequency $\ome$ (and their complex conjugates):
\begin{eqnarray}\label{eq:corrbathDenis}
&&\int\limits_0^\infty\mathrm{d}\tau\,e^{i\ome\tau}
\moy{ \hat{B}_\alpha^\dagger(0)\, \hat{B}_\beta(\tau)}_{\rm B}
=\delta_{\alpha\beta}\sum_k\frac{i|g_{\alpha{k}}|^2n_{\alpha{k}}}{\ome-\Omega_k+i0^+}
\equiv\delta_{\alpha\beta}[\gamma_\alpha\bar{n}_\alpha+i\delta_\alpha],\\
&&\int\limits_0^\infty\mathrm{d}\tau\,e^{i\ome\tau}
\moy{ \hat{B}_\alpha(\tau)\, \hat{B}_\beta^\dagger(0)}_{\rm B}
=\delta_{\alpha\beta}\sum_k
\frac{i|g_{\alpha{k}}|^2(n_{\alpha{k}}+1)}{\ome-\Omega_k+i0^+}
\equiv\delta_{\alpha\beta}[\gamma_\alpha(\bar{n}_\alpha+1)+i\delta_\alpha'],\label{eq:corrbathDenis2}
\end{eqnarray}
$n_{\alpha k}=(e^{\b\Omega_k}-1)^{-1}$ and $\bar{n}_\a=(e^{\b\ome}-1)^{-1}$ are Bose-Einstein distributions, 
$\g_\a\propto\restriction{|g_{\a k}|^2}{k\leftrightarrow\ome}$ and the Lamb-shifts $\d_\a,\d'_\a$ can be expressed through Cauchy principal values. The resulting master equation has the GKSL form
\begin{equation}
\begin{split}
&\dot\rho=
-i\left[ H_{\rm S}+H_{\rm LS}+ H_{\rm MW}, \rho\right]
+\mathcal{D}_1( \rho)+\mathcal{D}_2( \rho)\\
H_{\rm LS} \equiv \sum_{\a=1,2}\d'_\a\ketbra{\a}{\a}&-\d_\a\ketbra{0}{0}\quad\qquad
\mathcal{D}_\alpha( \rho)\equiv
L_\a\r L_\a^\dagger+ \wt L_\a\r \wt L_\a^\dagger-\frac12\arga{L_\a^\dagger L_\a+\wt L_\a^\dagger \wt L_\a,\r } \\
L_\a=&\sqrt{2\g_\a\bar n_\a}\,\ketbra{\a}{0}\quad\qquad
\wt L_\a=\sqrt{2\g_\a(\bar n_\a+1)}\,\ketbra{0}{\a}
\end{split}
\end{equation}

Let us make here a few comments on this GKSL equation to make contact with our DNP model:
\begin{enumerate}
 \item This equation is identical to the one obtained through Ref.~\cite{NR20}'s approach employed in the DNP model. 
 Indeed one can project the usual Born-Markov equation~\eqref{eq:markov} on eigenstates and get the same jump operators for 
 $\D,J\ll\ome$. Alternatively, 
 one can directly compute Ref.~\cite{NR20}'s nonsecular jump operators (defined in Eq.(6) of the main text) by redefining slightly the system 
 in order to satisfy the hypothesis that $H_{\rm int}$ is written as a sum of Hermitian operators : 
 $H_{\rm int}=\sum_{\a=1,2} \hat\SS_\a\otimes\hat \BB_\a$ with $\hat\SS_\a=\ketbra\a0 +\mathrm{H.c.}$, $\hat\BB_\a=\sum_k g_{\a k}\hat b_{\a k}+\mathrm{H.c.}$, yielding the same result.
 \item The toy model corresponds to a variant of our DNP model for $N=2$ spins truncated to the three lowest levels, \ie taking $\ket{++}=0$ or density matrix elements $\bra{\l}\r\ket{++}$ for any vector $\ket \l$. 
 More precisely, the relationship is $U_{12}\equiv J$ with fixed local disorder $\D_i\equiv \pm 2\D$ and energy levels shifted so that $H_{\rm S}\ket{--}=0$, only the spin 1 is irradiated at frequency $\mw$ (with $V\equiv\om_1/\sqrt2$) and 
 the bath do not couple to $S_i^z$ (\ie $T^*,T(0)\to\io$); for example one can consider $H_{\rm int}=\sum_i S_i^x\otimes B_i^x$ 
 which defines the parameters $\g_{1,2}\equiv\g(\ome)/8$. 
 In this setting, the populations of levels 1 and 2 in the 3-level model plotted in Fig.~\ref{fig:3levels}'s panels give directly the steady-state polarization of each spin: $\Tr(S_i^z\r)\equiv\r_{ii}-\frac12$.
\end{enumerate}

In the following we neglect the Lamb-shifts~$\delta$. Explicitly in components, \begin{eqnarray}
&&\dot{\rho}_{11}=iV^*e^{i\mw{t}}\rho_{01}^*-iVe^{-i\mw{t}}\rho_{01}
+iJ(\rho_{12}-\rho_{12}^*)+2\gamma_1\bar{n}_1\rho_{00}
-2\gamma_1(\bar{n}_1+1)\rho_{11},\qquad\\
&&\dot{\rho}_{22}=iJ(\rho_{12}^*-\rho_{12})+2\gamma_2\bar{n}_2\rho_{00}-2\gamma_2(\bar{n}_2+1)\rho_{22},\\
&&\dot{\rho}_{01}=i\ome\rho_{01}+iV^*e^{i\mw{t}}(\rho_{00}-\rho_{11})
-i\Delta\rho_{01}+iJ\rho_{02}-[\gamma_1(2\bar{n}_1+1)+\gamma_2\bar{n}_2]\rho_{01},\quad\\
&&\dot{\rho}_{02}=i\ome\rho_{02}-iV^*e^{i\mw{t}}\rho_{12}
+iJ\rho_{01}+i\Delta\rho_{02}-[\gamma_2(2\bar{n}_2+1)+\gamma_1\bar{n}_1]\rho_{02},\\
&&\dot{\rho}_{12}=2i\Delta\rho_{12}-iVe^{-i\mw{t}}\rho_{02}+iJ(\rho_{11}-\rho_{22})-[\gamma_1(\bar{n}_1+1)+\gamma_2(\bar{n}_2+1)]\rho_{12}.
\end{eqnarray}
Looking for the solution with time-independent $\rho_{11},\rho_{22},\rho_{12}$ and $\rho_{01},\rho_{02}\propto{e}^{i\mw{t}}$ (we rename $\rho_{0\a}\to\rho_{0\a}{e}^{-i\mw{t}}$), we arrive at a time-independent linear system. Let us for simplicity take $V=V^*$,  $\gamma_1=\gamma_2\equiv\gamma$, and set $\bar{n}_\alpha=0$ (\ie zero temperature). Then the system has the form
\begin{eqnarray}
&&0=V(\rho_{01}-\rho_{01}^*)-J(\rho_{12}-\rho_{12}^*)-2i\gamma\rho_{11},\\
&&0=J(\rho_{12}-\rho_{12}^*)-2i\gamma\rho_{22},\\
&&0=(\mw-\ome+\Delta-i\gamma)\rho_{01}
-J\rho_{02}+V(2\rho_{11}+\rho_{22}-1),\\
&&0=-J\rho_{01}+(\mw-\ome-\Delta-i\gamma)\rho_{02}+V\rho_{12},\\
&&0=V\rho_{02}-2(\Delta+i\gamma)\rho_{12}+J(\rho_{22}-\rho_{11}).
\end{eqnarray}
Let us solve the last three equations (we define $\D\om=\ome-\mw$ for compactness):
\begin{equation}
\begin{array}{l}
\rho_{01}D=-J^2V(\rho_{11}-\rho_{22})+[V^3+2V(\Delta+i\gamma)(\D\om+\Delta+i\gamma)](1-2\rho_{11}-\rho_{22}),\\
\rho_{02}D=-JV(\Delta-\D\om-i\gamma)(\rho_{11}-\rho_{22})
-2JV(\Delta+i\gamma)(1-2\rho_{11}-\rho_{22}),\\
\rho_{12}D=[J(\D\om+\Delta+i\gamma)(\D\om-\Delta+i\gamma)-J^3](\rho_{11}-\rho_{22})-JV^2(1-2\rho_{11}-\rho_{22}),\\
D=2J^2(\Delta+i\gamma)+V^2(\D\om-\Delta+i\gamma)-2(\Delta+i\gamma)(\D\om+\Delta+i\gamma)(\D\om-\Delta+i\gamma).
\end{array}
\end{equation}
In the linear response regime, we seek  $\rho_{01},\rho_{02}=O(V)$, $\rho_{12},\rho_{11},\rho_{22}=O(V^2)$ and neglect $V^2$ in the denominator, which gives
\begin{equation}
\rho_{11}=\frac{V^2[(\D\om+\Delta)^2+\gamma^2]}{(\D\om^2-\Delta^2-J^2-\gamma^2)^2+4\gamma^2\D\om^2},\quad
\rho_{22}=\frac{V^2J^2}{(\D\om^2-\Delta^2-J^2-\gamma^2)^2+4\gamma^2\D\om^2}.
\end{equation}
This is the expression plotted in Fig.~\ref{fig:3levels}'s panels.
We have good mixing at $\gamma,\Delta\ll{J}$, Anderson localization for $\gamma\ll{J}\ll\Delta$, Zeno localization at $\Delta<J\ll\gamma$, and weak incoherent coupling for $J\ll\gamma,\Delta$.

\newpage

\section{Microscopic model of the electronic spins and bath}\label{sec:bath}

In this section we comment on the choice of the microscopic Hamiltonian of the whole system. 
We discuss especially the coupling to the bath modes. We model the dissipation as coming from the interaction with vibrations of the glassy material in which the electronic spins are embedded. 
Our nonsecular equation assumes uncorrelated bath degrees of freedom. This may be brought about by two qualitatively different types of bath vibrational modes at low temperature (see Refs.~\cite{MSI2017,WNGBSF19} and references therein):
\begin{itemize}
 \item \underline{Spatially-localized vibrations:} the amorphous medium is a strongly disordered matrix which hosts localized or quasi-localized vibrations. These modes decay into the surrounding, and are more natural to describe in the coordinate basis than in the Bloch wave basis. As the electron spins are dilute, a vibration on a given spin cannot affect the local environment of other spins. Consequently they are decorrelated degrees of freedom.
\item \underline{Delocalized vibrations:} the glassy matrix supports as well delocalized vibrations (e.g. acoustic phonons or delocalized glassy modes) that are extended in the whole volume. They thus seem at odd with the previous argument: how can system-spanning vibrations act like a local bath for each spin? In the following we investigate this question within a simple model of ballistic phonons. We show that if a system is dilute enough, modes of large enough frequency (\ie short enough wavelength) are effectively uncorrelated. However the present treatment of the dissipation modes is somewhat simplistic, nevertheless it shows a microscopic example with bath rates that are power-law increasing with temperature, as expected from NMR experiments~\cite{FCSMRTC14,FECSRTCC16}.
\end{itemize}


\subsection{Coupling between spins and the magnetic field}\label{sec:coupB}

In absence of lattice vibrations, the electron spins are frozen in an amorphous matrix. We denote the position of a given spin by $\bm R $. 
The spin-orbit interaction between this spin and the external magnetic field depends on the orientation of the radical with respect to the field, and in general 
is thus written as a tensorial\footnote{Repeated Greek indices are summed over in the whole section. Note that the magnetic field is along $z$, hence the $z$ index.} coupling $-g_{z\m}\m_{\rm B}B S_i^\m$~\cite[Chap. VI]{abragam}-\cite{CCB08} where
\begin{equation}\label{eq:gtensor1}
  g_{z\m}=g_{\rm e}\d_{z\m}+g^{(0)}_{z\m}
\end{equation}
The first term in Eq.~\eqref{eq:gtensor1} is the dominant isotropic contribution, resulting in the Zeeman part of the spin Hamiltonian $H_{\rm S}$, \ie $ \sum_i \om_{\rm e}S_i^z=\ome S^z$
with $\om_{\rm e}=-g_{\rm e}\m_{\rm B}B$ the Zeeman frequency, related to the electron Land\'e $g$-factor $g_{\rm e}$ $(\simeq -2)$ through $\m_{\rm B}=\frac{e}{2m_{\rm e}}$. 
$e$ is the unit charge, $m_{\rm e}$ the electron mass and $\m_{\rm B}$ the Bohr magneton. 
In practice we take $\om_{\rm e}= 93.9\cdot2\p$GHz, meaning $B\simeq 3.35$ Tesla, a standard value for DNP. 
The second term in Eq.~\eqref{eq:gtensor1} is the so-called $g$-factor anisotropy of the disordered sample, which depends on the radical orientation and therefore appears as a random quantity. 
This term contributes to $H_{\rm S}$ as $ \sum_i \D_iS_i^z$ 
with\footnote{For notational simplicity we omit explicit reference to the spin $i$, implicitly born by $g^{(0)}_{zz}$.} $\D=-g^{(0)}_{zz}\m_{\rm B}B$.
Note that we discarded directions $\m\neq z$: when the magnetic field is large one can resort to the \textit{secular} approximation of the Hamiltonian, which consists in keeping only the terms conserving the total polarization along $z$. The reason in that hybridization between different polarization sectors is very weak in perturbation theory~\cite[Chap. IV.II.A]{abragam}-\cite{SPG92,DLRAMR16}. 
This rationale holds for the dipolar couplings as well, implying Eq. (2) of the main text. 
The last dipolar term $\propto U_{ij} S_i^z S_j^z$ has been dropped in the numerics as it commutes with all operators $ S_i^z$ and has no impact on the physics.

\subsection{Vibrational modes of the embedding material}

The position of an  electron spin in the amorphous matrix is $\bm r=\bm R +\bm u(\bm R)$ with 
$\bm R$ an equilibrium position,
and  $\bm u(\bm R)$ describes a small vibrational motion around it. The latter motion affects all space-dependent quantities, such as the Zeeman interaction
which is the strongest term in the Hamiltonian. 
The field $\bm u$ is expected to vary slowly on the scale of the electron distances for extended vibrational modes. The distance between two particles 1 and 2 in the matrix is indeed $\bm r=\bm r_1-\bm r_2=\bm R_1-\bm R_2+ (\partial_\m\bm u) (\bm R_1-\bm R_2)_\m$ at first order in the $\bm u$ derivatives.
The tensorial coupling in Eq.~\eqref{eq:gtensor1} is now modified by the vibrations as~\cite[Chap. 22]{AM}-\cite{landauVII}.
\begin{equation}\label{eq:gtensor}
  g_{z\m}=g_{\rm e}\d_{z\m}+g^{(0)}_{z\m}+g^{(1)}_{z\m\g\d}\partial_\g u^\d+ g^{(2)}_{z\m\g\d\g'\d'}\partial_\g u^\d \partial_{\g'} u^{\d'}+\dots
\end{equation}
Vibrational modes in the glass originate from several processes~\cite{MSI2017,WNGBSF19}, but as mentioned in the introduction of this section, in the following we shall focus on extended modes, as the case of localized vibration modes satisfies more intuitively our assumption of spatially-uncorrelated degrees of freedom. For simplicity we model delocalized vibrations as low-energy excitations arising from acoustic phonon modes.
In the glassy sample, the arrangement of the different atoms is not periodic. This implies a continuous set of wavevectors; we nonetheless use the standard theory of phonons of a periodic lattice for convenience, as it should not affect much the results.  
One can write the quantized displacement field~\cite{AM}
\begin{equation}\label{eq:vib}
 \bm u(R)=\frac{1}{\sqrt N}\sum_{\bm k,s}\frac{\bm{e}_{\bm{k}/k,s}}{\sqrt{2m\Omega_{\bm k,s}}} \, a_{\bm k,s}\, e^{i\bm k\cdot\bm R}\quad+ \textrm{H.c.}
\end{equation}
$s=1,2,3$ is the polarization index, $\bm k$ are the wavevector (quantized in the first Brillouin zone due to the assumed periodicity of the lattice), $m$ is the mass of the glassy molecule\footnote{\ie approximating pyruvic acid as a monoatomic substance.}
$\Omega_{\bm k,s}$ are the phonon frequencies, $\bm{e}_{\bm{k}/k,s}$ are polarization unit eigenvectors, 
and $a_{\bm k,s}$, $a^\dagger_{\bm k,s}$ are phonon annihilation and creation operators.
The bath Hamiltonian is thus a collection of harmonic oscillators $H_{\rm B}=\sum_{\bm k,s}\Omega_{\bm k,s}a^\dagger_{\bm k,s}a_{\bm k,s}$.

In the following we consider separately the one- and two-phonon processes as they are incoherent  owing to Wick's theorem~\cite{coleman}, \ie their contribution to the correlation function $\moy{B_i^\m(t)B_i^\m}_{\rm B}$ is additive.

 \subsubsection{Direct process}\label{sub:direct}

The direct process concerns the exchange of a single phonon between the bath and the system. It is due to the first-order interaction 
between the spin and the bath modes in Eq.~\eqref{eq:gtensor} involving the tensor $g^{(1)}$, substituting $\partial_\g u^\d$ via Eq.~\eqref{eq:vib}. The interaction Hamiltonian is thus of the form~\eqref{eq:Hint} where $B_i^\m$ a linear combination of annihilation and creation operators. 
The equilibrium bath correlation function~\eqref{eq:defReIm} then involves only quadratic correlators in the $a_{\bm k,s}$.
For simplicity we drop tensor indices and the polarization vectors, as these factors only contribute $O(1)$ proportionality constants. 
The calculation is standard (as done in~Eqs.\eqref{eq:corrbathDenis}-\eqref{eq:corrbathDenis2}) using the dispersion relation 
\begin{equation}
 \Omega_k=kv
\end{equation}
with $v$ the sound velocity. In the continuous limit $N\to\io$ we replace the sums over wavevectors with integrals 
\begin{equation}\label{eq:sumcont}
\begin{split}
\sum_{\bm k,s}\,\bullet \ \longrightarrow&\  \argp{\frac{L}{2\p}}^3\int\dd \bm k\, \bullet=\argp{\frac{L}{2\p}}^3\int \dd\Omega\, 4\p\argp{\frac{\Omega}{v}}^2\frac1v\,\bullet
\end{split}
\end{equation}
and get the bath rate
\begin{equation}\label{eq:direct}
\frac{1}{T(|\om|)}=\frac{1}{2\p}\argp{\frac{g^{(1)}}{g_{\rm e}}}^2\frac{\om_{\rm e}^2}{\r_0 v^5}\om^3\coth\argp{\frac{\b\om}{2}}
\end{equation}
where $\r_0=Nm/L^3$ is the mass density.

\subsubsection{Two-phonon processes}\label{sub:Raman}

The next term in the system-bath interaction is a two-phonon process caused by the second-order interaction between the spin and the bath modes in Eq.~\eqref{eq:gtensor} involving the tensor $g^{(2)}$.  
It takes the form~\eqref{eq:Hint} where $B_i^\m$ is quadratic in the annihilation and creation operators: 
\begin{equation}\label{eq:defHBRaman}
\begin{split}
  B_i^\m=\frac{\om_{\rm e}}{2mN}\frac{g^{(2)}_{z\m\g\d\g'\d'}}{g_{\rm e}}\argp{\sum_{\bm k,s} c^{\g\d}_{\bm k,s}a_{\bm k,s} +c^{\g\d *}_{\bm k,s}a^\dagger_{\bm k,s}}\argp{\sum_{\bm k',s'} c^{\g'\d'}_{\bm k',s'}a_{\bm k',s'}  +c^{\g'\d'*}_{\bm k',s'}a^\dagger_{\bm k',s'}}\quad\textrm{with}\quad
 c^{\g\d}_{\bm k,s}=\frac{k^\g e^\d_{\bm{\hat k},s}}{\sqrt{\Omega_{\bm k,s}}} e^{i\bm k\cdot\bm R_i}
\end{split}
 \end{equation}
 For notational convenience, we here drop some indices such as the polarization $s$ and some factors which will be reinstated later on. Sums over indices are implicit. We thus have to deal with 4-point canonical averages applying Wick's theorem~\cite{coleman}, for example
$$ \moy{a_{\bm k_1} a_{\bm k_2}^\dagger a_{\bm k_3}a^\dagger_{\bm k_4}}_{\rm B}
=\d_{\bm k_1,\bm k_2}\d_{\bm k_3,\bm k_4}(n_{\bm k_1}+1)(n_{\bm k_3}+1)+\d_{\bm k_1,\bm k_4}\d_{\bm k_2,\bm k_3}(n_{\bm k_1}+1)n_{\bm k_2}$$
with $n_{\bm k}$ the Bose-Einstein factor at frequency $\Omega_{\bm k}$. 

We get for the bath correlations
\begin{equation}
\begin{split}
  \moy{B(t)B(0)}_{\rm B}= &A e^{-i(\Omega_{\bm k_1}+\Omega_{\bm k_2})t}+Be^{i(\Omega_{\bm k_1}+\Omega_{\bm k_2})t}
 +Ce^{i(\Omega_{\bm k_1}-\Omega_{\bm k_2})t}+De^{-i(\Omega_{\bm k_1}-\Omega_{\bm k_2})t}\\\
 A=&\moy{a_{\bm k_1} a_{\bm k_2} a_{\bm k_3}^\dagger a^\dagger_{\bm k_4}}_{\rm B}c_{\bm k_1} c_{\bm k_2} c_{\bm k_3}^* c^*_{\bm k_4}\\
 B=&\moy{ a_{\bm k_1}^\dagger a^\dagger_{\bm k_2}a_{\bm k_3} a_{\bm k_4}}_{\rm B}c_{\bm k_1}^* c_{\bm k_2}^* c_{\bm k_3} c_{\bm k_4}\\
 C=&\moy{ a_{\bm k_1}^\dagger a_{\bm k_2}a_{\bm k_3} a^\dagger_{\bm k_4}}_{\rm B}c_{\bm k_1}^* c_{\bm k_2} c_{\bm k_3} c_{\bm k_4}^*
 +\moy{ a_{\bm k_1}^\dagger a_{\bm k_2}a_{\bm k_3}^\dagger a_{\bm k_4}}_{\rm B}c_{\bm k_1}^* c_{\bm k_2} c_{\bm k_3}^* c_{\bm k_4}\\
 D=&\moy{a_{\bm k_1} a_{\bm k_2}^\dagger a_{\bm k_3} a^\dagger_{\bm k_4}}_{\rm B}c_{\bm k_1} c_{\bm k_2}^*  c_{\bm k_3}c^*_{\bm k_4}
 +\moy{a_{\bm k_1} a_{\bm k_2}^\dagger a_{\bm k_3}^\dagger a_{\bm k_4}}_{\rm B}c_{\bm k_1} c_{\bm k_2}^* c_{\bm k_3}^* c_{\bm k_4}
\end{split}
\end{equation}
which then give, taking the real part of the Fourier transform $\G(\om)$~\eqref{eq:corrfbath},\eqref{eq:defReIm}:
\begin{equation}
 \frac{\g(\om)}{2}=\underbrace{A\p\d(\om-\Omega_{\bm k_1}-\Omega_{\bm k_2})}_{\textrm{absorption of 2 phonons}}+\underbrace{B\p\d(\om+\Omega_{\bm k_1}+\Omega_{\bm k_2})}_{\textrm{emission of 2 phonons}}
 +\underbrace{C\p\d(\om+\Omega_{\bm k_1}-\Omega_{\bm k_2})+D\p\d(\om-\Omega_{\bm k_1}+\Omega_{\bm k_2})}_{\textrm{Raman process: absorption and emission}}
\end{equation}
We now reinstate some factors from the definition of $B_i^\m$ and $c^{\g\d}_{\bm k,s}$ in Eq.~\eqref{eq:defHBRaman}, 
and we replace the sums over wavevectors with integrals as in Eq.~\eqref{eq:sumcont}. 
We shall for convenience with the delta functions apply the following formulas:
\begin{equation}
  n_{k}= n(\Omega_k)=\frac{1}{e^{\b\Omega_k}-1}=\frac{e^{-\b\Omega_k/2}}{2\sinh(\b\Omega_k/2)}\,,\qquad
  n_{k}+1=\frac{e^{\b\Omega_k/2}}{2\sinh(\b\Omega_k/2)}
\end{equation}
In the continuous limit for wavevectors and at low temperature\footnote{The Debye frequency $\Omega_{\rm D}$ must be large, \ie $\b\Omega_{\rm D}\gg1$, which is usually the case around $1$ K.}, dropping again indices and polarization vectors, we obtain 
\begin{equation}
\begin{split}
  \g(\om)=&2\om_{\rm e}^2\argp{\frac{g^{(2)}}{g_{\rm e}}}^2e^{\b\om/2}\frac\p8\int\dd k\dd k'\,\frac{4\p k^3}{(2\p)^3\r_0v\sinh(\b vk/2)}\frac{4\p k'^3}{(2\p)^3\r_0v\sinh(\b vk'/2)}\\
  &\hskip150pt \times \argc{\d(\om-vk-vk')+\d(\om+vk+vk')+2\d(\om-vk+vk')}
\end{split}
\end{equation}
which translates into the bath rate~\eqref{eq:detbal}:
\begin{equation}\label{eq:raman}
 \frac{1}{T(|\om|)}=\frac{16}{\p^3}\argp{\frac{g^{(2)}}{g_{\rm e}}}^2 \frac{\om_{\rm e}^2}{(\r_0 v^5)^2}
 T^7\cosh\argp{\frac{\b\om}{2}}
 \argc{\underbrace{\int_0^{\b|\om|/2}\dd y\, \frac{y^3\argp{\frac{\b|\om|}{2}-y}^3}{\sinh y\sinh\argp{\frac{\b|\om|}{2}-y}}}_{\textrm{absorption of two phonons}}+\underbrace{2\int_0^{\io}\dd y\, \frac{y^3\argp{\frac{\b|\om|}{2}+y}^3}{\sinh y\sinh\argp{\frac{\b|\om|}{2}+y}}}_{\textrm{Raman process: absorption and emission}}}
\end{equation}
In order to discuss the limit $\b\abs{\om}\ll1$ in the next section, the two-phonon absorption integral can be rewritten as
\begin{equation}\label{eq:ramanlowbetaomega}
\argp{\frac{\b|\om|}{2}}^7 \int_0^{1}\dd z\, \frac{z^3\argp{1-z}^3}{\sinh \argp{\frac{\b|\om|}{2}z}\sinh\argc{\frac{\b|\om|}{2}\argp{1-z}}}
\end{equation}

\subsubsection{Discussion of the one- and two-phonon bath timescales}

In the considered range of temperatures here, $T$ goes roughly from 25$\cdot2\p$GHz to 250$\cdot2\p$GHz. The $S_i^z$ flips have energy much lower, in the MHz range, and $S_i^x$, $S_i^y$ flips have energy around $\ome=93.9\cdot2\p$GHz. 
For our range of frequencies and temperature, the correct limit to consider is thus $\b\abs{\om}\ll1$. \\
In this limit, the direct rate~\eqref{eq:direct} is $\propto \om^2 T$, it increases linearly with temperature. 
The Raman process contribution, from Eq.~\eqref{eq:raman}, has a rate $\propto T^7$ instead and is roughly independent on frequency. Linearizing both hyperbolic sines in this limit in the integrand of~\eqref{eq:ramanlowbetaomega}, we see that the two-phonon absorption contribution is $\propto\abs{\om}^5 T^2$, \ie quadratic in temperature.

In experiments the dependence in frequency is not known while the dependence in temperature of $T_1=2T(\ome)$ is roughly $T^{2}$~\cite{FCSMRTC14,FECSRTCC16}, as in the two-phonon absorption. As we do not know realistic values of the ratios $g^{(i)}/g_{\rm e}$, we cannot assess  the order of magnitude of the predicted timescales, 
as well as the relative weight of the different rates between them,  which determines the temperature dependence of the bath correlation function $\g(\om)$. 
In any case we note that all of the rates in this model increase with temperature, a property that is crucial to the Zeno physics discovered in this work.

For a recent review of spin-lattice relaxation, we refer the reader to Ref.~\cite{Lu23}.

\subsection{Decorrelation of the bath degrees of freedom}

Finally let us look at the decorrelation assumption of the bath degrees of freedom. 
The fact that ${\moy{B_i^\m(t)B_j^\n}_{\rm B}\propto\d_{\m\n}}$ owes to the isotropy of the material. 
If $i\neq j$ the only difference in the calculation is that all integrals over wavevectors (say $\bm k$) get an additional phase factor 
$e^{i\bm k\cdot (\bm R_i-\bm R_j)}$. Therefore decorrelation happens if the wavelength of the phonon is
much smaller than the inter-electron distance $|\bm R_i-\bm R_j|$. In other words, the criterion for decorrelation is that the phonon frequencies $\om$ involved 
are such that $\om\gg v/|\bm R_i-\bm R_j|$ \ie $\om\gg v \r_{\rm e}^{1/3}$ where $\r_{\rm e}$ is the electron density in the material. In practice ($\r_{\rm e}\sim 10$ mmol/L, $v\sim 10^3\ \textrm{m}/ \textrm{s}$) this threshold is of the order of magnitude of the $2\p$GHz. 
This mechanism of spatial decorrelation of delocalized modes is general, but this simplistic model of ballistic phonons may well not give a realistic description. Nevertheless it represents a simple example with a (power-law) decrease of the bath relaxation timescales, as generally expected and displayed by $T_1$ measurements, needed to Zeno localize as temperature is raised, as shown in the main text. 

\end{widetext}

\bibliographystyle{apsrev4-1.bst}
\bibliography{MBL}

\end{document}